# Migration processes in the Solar System and their role in the evolution of the Earth and planets

M Ya Marov, S I Ipatov

## Contents

**Abstract.** We discuss problems of planetesimal migration in the emerging Solar System and exoplanetary systems. Protoplanetary disk evolution models and the formation of planets are considered. The formation of the Moon and of the asteroid and trans-Neptunian belts is studied. We show that Earth and Venus could acquire more than half of their mass in 5 million years, and their outer layers could accumulate the same material from different parts of the feeding zone of these planets. The migration of small bodies toward the terrestrial planets from various regions of the Solar System is simulated numerically. Based on these computations, we conclude that the mass of water delivered to the Earth by planetesimals, comets, and carbonaceous chondrite asteroids from beyond the ice line could be comparable to the mass of Earth's oceans. The processes of dust migration in the Solar System and sources of the zodiacal cloud are considered.

M Ya Marov (*), S I Ipatov (**)
V.I. Vernadsky Institute of Geochemistry and Analytical Chemistry,
Russian Academy of Sciences,
ul. Kosygina 19, 119991 Moscow, Russian Federation
E-mail: (*) marovmail@yandex.ru, (**) siipatov@hotmail.com,
(**) ipatov@geokhi.ru

## 1. Introduction

The migration of bodies is one of the key processes in the formation and evolution of a planetary system. Migration processes make a significant contribution to the transfer of planetesimals and to the dynamics of the resulting configurations (planetary system architecture) at the formation stage of planet embryos and their satellite systems, and retain their role throughout the evolution with regard to the most dynamical small bodies (asteroids, comets, and meteoroids). Due to migration, matter was transferred during the formation of Solar System planets from the regions of the giant planets and the Kuiper belt toward the terrestrial planets, with a fundamental impact on their nature. In the modern era, the asteroid–comet hazard (ACH) problem for the Earth is directly related to the migration of small bodies. Planet formation models and the fallout of planetesimals on growing planets, initially formed at different distances from the Sun, have been considered in numerous studies by researchers in Russia and elsewhere, including the authors of this paper, by numerically solving the relevant physical and dynamical problems of planetary cosmogony. The structure and formation of the Solar System, including the data on planets, their satellites, and small bodies, are considered in detail in [1–3].

In this paper, we discuss the problems of planetesimal migration and planetary dynamics in the emerging Solar

System, including models of the migration of bodies from different regions of the Solar System toward the terrestrial planets. The formation of planetary bodies and asteroid and trans-Neptunian belts is considered. The migration of dust in the Solar System and the sources of the zodiacal cloud are studied. The problems of migration of planetesimals and planets in exoplanetary systems are discussed based on similarities with the dynamical properties of the Solar System. The original models considered by the authors serve to develop modern concepts in key areas of stellar and planetary cosmogony.

## 2. Formation of the protoplanetary disk and planetesimals

### 2.1 Formation and evolution of the protoplanetary disk

Stars are born in gradually compressing clusters during the fragmentation of interstellar clouds with the formation of a protostellar nebula, the cradle of a star and its planetary system. According to currently accepted notions, the sequence of planetary system formation processes includes the formation of an accretionary gas–dust disk around the parent star (mainly of the late spectral type) and its decay into primary clumps, from which solid bodies (planetesimals), planetary embryos, and eventually the planets themselves form. The key role in this sequence is played by the different (hydrodynamic and gravitational) instability types in the disk, which initiate its fragmentation, the accretion of solids and their subsequent growth, and various dynamical processes, among which the leading role is played by resonances, tidal effects, and migration of planetesimals and planet embryos (see, e.g., [4–7]).

A significant contribution to the formation of a protosolar nebula could be made by the explosion of a relatively nearby supernova and the appearance of a shock front. This would lead to additional cloud compression and implantation of short-lived isotopes such as $^{26}$Al and $^{60}$Fe, which have a significant impact on the cloud heating and evolution at an early stage [8, 9]. Evidence of such a process is the enrichment of a number of meteorites with the stable $^{26}$Mg isotope, which is produced from the parent isotope $^{26}$Al with a half-life of $\sim 0.72$ Myr [10], of which the daughter $^{60}$Fe isotopes are much more short-lived.

The cloud contains only 1 to 2% of dust particles by mass. Due to the rapid rotation, it condenses and flattens, which causes the formation of a hot dense thickening in the central region (a proto-Sun overcoming the threshold of thermonuclear fusion reactions), surrounded by a gas–dust accretion disk made of the remaining initial material of the nebula, whose mass is no more than 2 to 3% of the proto-Sun mass according to estimates. This disk serves as a cradle for the formation of planets and small Solar System bodies and is obviously analogous to protoplanetary circumstellar disks. The subsequent evolution includes continued accretion of the nebula matter onto the disk and simultaneous partial accretion of the disk matter onto the proto-Sun until this process is superseded by an intensive sweeping out of gas and volatile components and the removal of high-temperature condensates by the radiation pressure and proto-Sun plasma from its vicinity to the periphery of the Solar System [2]. This is evidenced by the presence of refractory chondrules embedded in the chondrite matrix at radial distances $R > 2$–$3$ AU.

This scenario underlies modern cosmogonic models based on astronomical observations of circumstellar protoplanetary disks, their structural features, and numerous exoplanet systems of both single and binary stars (see, e.g., [11–17]). An enormous contribution to the study of circumstellar disks and planetary system formation was made by observations of their structure, composition, and dynamics obtained on millimeter waves by the network of ground-based radio telescopes ALMA (Atacama Large Millimeter Array); this includes studies within the Resolving Star Formation with ALMA program and Protostellar Interferometric Line Survey (PILS) (see [18–21]). Together with IR observations from the Hubble, Spitzer, and Herschel space telescopes, they produced a breathtaking picture of all the protostellar nebula components combining to create planetary systems out of this 'cosmic stew.'

Reconstruction of the formation and growth of primary solid particles in a gas–dust protoplanetary disk at an early stage of evolution is an extremely difficult task, which can only be properly defined and solved using mathematical modeling methods, with a number of constraints imposed by the available results of astronomical observations and laboratory experiments on modeling particle interactions. A protoplanetary disk is a gas–dust turbulent medium with a magnetic field, which generally requires the use of heterogeneous mechanics and magnetohydrodynamic methods. Setting aside plasma effects, the motion of the gas–dust medium in the disk is modeled most properly in the framework of the mechanics of heterogeneous turbulent media with numerous factors taken into account, including the physicochemical properties of phases, heat and mass transfer, variations in the medium opacity to stellar radiation, viscosity, chemical reactions, phase transitions (position of the evaporation–condensation border), and coagulation. A rigorous mathematical treatment of these problems was given in monographs [22, 23] based on numerous publications by the authors on planetary cosmogony problems. They analyze the nature of the dynamical interaction between turbulent gas and dust, including the effect of the turbulence energy of the carrier phase on the behavior of solid particles and the inverse effect exerted by the dust component on the dynamical and thermal regimes of the gas phase, with the coagulation processes occurring in the disk taken into account. According to modern concepts, the collapsing gaseous shell of a nebula loses significant mass in the active young sun state at the T Tauri stage, before it joins the main sequence of the H–R diagram, with a characteristic time $\sim 10$ Myr. The loss of gas terminates much later, at the stage of the growth of dust particles to the initial solids, and therefore the interaction of particles with gas must be taken into account in modeling.

### 2.2 Formation and evolution of dust clusters

Numerical experiments have shown that it is not individual particles but their agglomerations — dust clusters — that merge much more easily in mutual collisions. According to [23, 24], rarefied dust clusters with a fractal structure and their interaction during collisions at moderate velocities are the key mechanisms for the agglomeration of dust particles and the growth of primary solids, which thus become a basis for the subsequent formation of planetesimals and planetary embryos. Physically, such a process seems to be substantiated quite well. The actual structure of dust clusters has an extremely complex and irregular geometry, and, although the mass fractal dimension used does not fully reflect the geometric properties of the fractal, it nevertheless allows taking the main properties of loose fractal structures into account when modeling cluster–cluster association processes [25–30] and spatiotemporal evolution. Not all particles of the gas–dust disk necessarily belong to clusters: some of them can form dust clouds that fill the Hill sphere and later fall onto planetesimals and planetary embryos.

Collisional interactions of dust clusters lead to the formation of denser structures, and the clusters themselves can contain both dense and loose (porous) particles [24, 31]. The clusters presumably also have a porous or fluffy structure, and, mimicking the tendency to form snow particles, are capable of forming very loose conglomerates of a fractal nature. This greatly facilitates the mathematical modeling of the growth of bodies in the disk due to the collision of clusters and particles inside them. When a large number of small dust clusters combine, homogeneous 'fuzzy' ('shaggy') aggregates form that have self-similar properties at short distances and whose inner voids gradually increase in volume, with a simultaneous increase in the average density of the merging bodies [24].

Indeed, such fluffy aggregates, due to their extremely high porosity, are resistant to destructive collisions at high impact velocities, and their radial drift in the disk is very slow. For typical shaggy aggregates of a relatively large geometric cross section compared to compact dust particles, the entire regime of motion in the gas carrier flow changes; in particular, the conditions for the occurrence of flow instability change due to significant changes in the aerodynamic drag of dust and gas. In addition, the efficiency of bouncing of colliding porous structures can change significantly [1, 24].

Thus, we regard the set of loose dust clusters of a protoplanetary subdisk as a special type of continuous (fractal) medium that has points and regions not filled with components, which greatly facilitates the process of merger/growth of particles and the formation of primary solids. We note that spectral observations of disks around young T Tauri stars indicate that fine ($\lesssim 1\,\mu$m) dust particles persist in them for 1–10 Myr. At the same time, in accordance with model estimates, much larger clumps of the same mass could grow during this time in the inner part of the disk (at distances $R < 10$ AU from the star),

Obviously, each of the proposed mechanisms of formation of large solid bodies in the disk with the subsequent formation of planetesimals and planetary embryos is based on the concept of the merger of the initial nanometer-size dust particles during their collisional interactions. But, as we have already mentioned, the direct merger of even small dust particles is ineffective, as is evidenced by estimates and results of laboratory experiments. Especially problematic as regards growth is the range from centimeter- to meter-size bodies. But even in the nanometer and micrometer-size ranges (the typical size of dust in interstellar clouds), the growth mechanism of solid dust particles, apparently driven mainly by van der Waals forces and electrostatic interactions, is problematic. Some exceptions may be provided by particles of amorphous water ice, concentrated beyond the ice line. Meanwhile, there is quite definitive evidence of the accretion of fairly large bodies such as pebble-cobblestones and their further association in the form of 'pebble piles' bound by gravitational forces.

### 2.3 Formation of rarefied clumps and planetesimals
Turbulence strongly affects the evolutionary processes in the disk. As turbulence phases out, the particles descend to the central plane of the disk, where they create a dusty subdisk, which in the course of its further compaction and the development of gravitational instability decomposes into primary clumps (dispersive dust clusters) of a wide range of sizes. Collisional interactions of the clumps and a progressive increase in particle size then gradually give rise to intermediate-size solid bodies (planetesimals) and planetary embryos.

Historically, several scenarios for such a process have been proposed. According to [32], the initial planetesimals are $\sim 100$ km in size in the zone of Neptune and $\sim 1$ km in the zone of the terrestrial planets. Greenberg et al. [33] supposed that most of the planetesimals in the Neptune zone were assumed to much smaller, of the order of a kilometer. According to estimates by Safronov and Vityazev [34], the initial masses of planetesimals were $\sim 10^{20}$ g in Earth's zone, reaching $\sim 10^{26}$ g in Jupiter's zone, $10^{27}$ g in Saturn's, and $3 \times 10^{27}$–$10^{28}$ g in Uranus's and Neptune's. The model by Eneev and Kozlov [35, 36] was fundamentally different: according to it, there was a certain mechanism for delaying the contraction of clumps, which combined into giant rarefied protoplanets about the size of their Hill spheres, and planets were formed during their subsequent contraction. In the 1990s, models that assumed the aggregation of planetesimals from small solids gained popularity (see, e.g., [37]), because self-generated turbulence in the dust subdisk was believed to prevent the accretion of dust clumps and the formation of protoplanetesimals due to gravitational instability.

In the early 2000s, new arguments were found to support the formation of rarefied dust clumps—clusters [38–61]. In addition to dust, such clumps could include decimeter-size objects such as pebbles and larger bodies (cobblestones) formed during collisions of dust clusters. In contrast to the planet formation region, in the trans-Neptunian region at radial distances greater than 30 AU from the Sun, shear turbulence in the dusty subdisk does not extend to the zone near the equatorial plane, where the critical density is reached more easily and the disk breaks up into dust clusters [38]. According to estimates, relatively small ($\sim 10^9$ cm) disk fragments could rapidly contract, forming bodies $\sim 10$ km in size and larger ($\sim 100$ km) planetesimals in about $\sim 10^6$ years. Studied in [60] was the formation of solid planetesimals with radii from 10 to several thousand kilometers during the contraction of clusters the size of a Hill sphere, located at about 40 AU from the Sun. The mechanisms of the formation and growth of dust clumps due to dust absorption were considered by Marov et al. [56] with the thermodynamic constraints taken into account, including the effect of dust on the opacity of the gas–dust medium and the dissipation of turbulent energy, with the conclusion that, in Earth's zone, such dense clumps (protoplanetesimals) with a mass of $\sim 10^{21}$–$10^{22}$ g and a radius of 50–100 km could form in $\sim 10^3$–$10^4$ years, depending on the initial particle sizes.

In [43], evidence was obtained for the effective formation of gravitationally bound clumps in a mass range corresponding to planetesimals 100 to 400 km in radius formed during contraction in the asteroid belt and 150 to 730 km in radius in the trans-Neptunian belt. Trans-Neptunian objects (TNOs) with semimajor axes 30 to 50 AU, often called the Kuiper belt or the Edgeworth–Kuiper belt (although, we note, G Kuiper, who studied this problem in 1951, believed that such a belt existed only during the formation of the Solar System and is currently nonexistent). The presence of this belt was predicted much earlier by other researchers: Frederick C Leonard and Armin O Leuschnera in 1930 and Kenneth Edgeworth in 1943. The masses of the formed planetesimals obtained in the computations in [43] are proportional to $R^{3/4}$, and therefore at a distance of 30 AU they are 5 to 6 times larger than in the asteroid belt. This is consistent with the characteristic maximum masses of asteroids and TNOs. The possibility of the formation of binary planetesimals was shown in [45]. Arguments in favor of the emergence of initially large asteroids (with diameter $d > 100$ km) were discussed in [50]; in [62], the size of TNOs was shown to not exceed 400 km, with larger objects slowly growing due to interaction with surrounding bodies. The results of observations of particle ejection from comets 67P and 103P [59, 63–65] testify in favor of the formation of comet nuclei during the contraction of rarefied clusters consisting of particles in the millimeter, centimeter, and decimeter size ranges.

An alternative (or indeed complementary) approach to the development of gravitational (Jeans) instability in the disk, which is responsible for the formation of initial clumps, planetesimals, and large solids, is of a hydrodynamic nature. It is also known as streaming instability. The main concept is the imbalance between the surface gas–dust density and mass transfer [66]. Two main scenarios for such instability have been proposed. The first is based on the idea that disk/subdisk turbulence can create localized regions with a high dust-to-gas ratio, which then grow and eventually reach the size of large bodies [67]. It is assumed that there is either a passive accumulation of particles by turbulence on large scales comparable to the dissipative interval of turbulence or accumulation of particles inside turbulent eddies, which, as it were, play the role of a trap, as was also noted in [23, 68]. The appearance of such formations in zonal flows [69], including aerodynamically emerging regions between vortices [46, 70], is possible. The second scenario assumes a feedback between gas and clumped particles in a two-phase flow, i.e., back reaction of particles on the gas flow. Such interaction between gas and dust is usually referred to as linear flow instability [67, 71]; it is responsible for the generation of initial protoplanetesimal embryos. Numerical simulation yields the dust-to-gas density ratio and some other parameters necessary for the implementation of such a mechanism. Large dust 'lumps,' especially those containing centimeter-size grains formed in a preceding collision and coagulation/coalescence processes, can have a significant impact on the stability of the flow of the gas–dust medium. It has also been shown that the nonlinear evolution of flow instability can be attended by gravitational instability at lower dust-to-gas ratios [39, 40, 52, 72–76]. In addition, shear turbulence caused by different gas and dust velocities in an inhomogeneous substance disk is responsible for the occurrence of the Kelvin–Helmholtz instability [23, 68, 77].

We note that, in the flow instability model, the motion of solid particles in a gas is considered in the case where the total mass of particles is not less than the mass of the gas (turbulence is suppressed at this mass ratio). The pressure gradient in the gas causes its orbital velocity to be less than the Keplerian velocity of particles at that distance from the Sun. Due to the difference in particle and gas velocities, particles lose angular momentum and spiral toward the Sun. The effect of gas drag leads to a spontaneous agglomeration of particles into gravitating clusters. Initially, small clusters increase the speed of the gas, move faster than individual particles, spiral more slowly toward the Sun, and thereby enhance the capture of drifting particles. In calculations, particle clusters form giant filament-like structures that can merge in collisions.

Having reached the density required for gravitational collapse (three orders of magnitude higher than the gas density), the filament-like structures collapse gravitationally and disintegrate into large clumps the size of a Hill sphere. In the calculations in [39], the mass of the largest cluster was 3.5 times the mass of Ceres. These clumps contract to form planetesimals. In the calculations in [45], the initial mass distribution of planetesimals formed during flow instability satisfies the power law $dn/dM \sim M^{-1.6}$. Flow instability works better beyond the ice line.

Flow instability is associated with dense clouds of particles ranging from a millimeter to a meter in size, formed in a turbulent gas–dust medium due to the rapidly increasing convergence of the radial drift and the difference between the velocities of the dust and gas phases [58]. This mechanism allows millimeter-size particles to form at distances $R > 10$ AU from the Sun, and its efficiency increases as the particle size increases, suggesting that it can be responsible for the formation of planetesimals from the collapse of clumps roughly 100–1000 km in size for both asteroids and TNOs. In flow instability calculations, the typical particle size in models [39, 41, 45] was from 10 cm to 1 m. Along with models with structurally compact and monodisperse dust particles, a model with polydisperse fractal particles formed in dust clusters as a result of coagulation has also been proposed [78].

We emphasize that, at a sufficiently high dust-to-gas ratio, turbulence can be suppressed and planetesimals can form only via gravitational (Jeans) instability, without the participation of flow instability. For a collection of loose dust clusters of a protoplanetary subdisk, viewed as a special type of continuous medium with a fractal structure, a modified gravitational instability criterion was proposed, borrowed from kinetic theory based on nonextensive Tsallis statistics and related generalized Euler hydrodynamic equations [31].

Along with gravitational and hydrodynamic instabilities, other models of the formation of rarefied clumps have been considered. For example, the formation of planetesimals due to turbulence caused by magnetorotational instability was studied in [42]. In this case, the masses of the formed planetesimals reached several ten Ceres masses. The instability caused by Rossby waves in a gas flow containing particles 1–100 cm in size was considered in [48]. The formation of dense clusters of millimeter-size particles was studied by Cuzzi et al. [46, 47], with the conclusion that some of them turn into 'sandpile' planetesimals like TNOs 10–100 km in diameter. In general, small bodies are dominant in the size distribution of planetesimals, although the mass is concentrated in several large bodies. Finally, we mention the model in [48, 49], where over 300 gravitationally bound nuclei were formed after 200 orbital periods, with 20 nuclei being more massive than Mars. Interestingly, such large embryos formed from initial clumps made of bodies the size of cobblestones in only 40 revolutions around the Sun.

Attempts have been made to obtain model estimates of the formation time of planetesimals and planetary embryos. A number of authors have presented arguments in favor of large protoplanetesimal contraction times. According to Vityazev et al. [79], in the feeding zone of the terrestrial planets, the clump contraction time is about $10^5$–$10^6$ years. According to Safronov [4], the clump contraction time is about $10^4$ years at 1 AU and $10^6$ years at Jupiter's separation from the Sun. The time of formation of TNOs up to 1000 km in diameter was estimated by Makalkin and Ziglina [38] to be several million years, with most of this time occupied by dust settling in the equatorial plane, while the clumps contracted in about 1 Myr with 1000 km objects formed by contraction of clusters containing 10 km bodies. It was shown by Myasnikov and Titarenko [80, 81] that the lifetime of such gas–dust clusters can exceed several million years, depending on the optical properties of the particles and the concentration of short-lived radionuclides.

There are alternative estimates, however. In [51], the time of planetesimal formation from rarefied clumps was shown to decrease with distance from the Sun. According to [43], it took 25 revolutions around the Sun for the maximum values of the particle mass per unit volume to become about 3000 times the gas density in the central plane. It was shown in [47] that some dense clumps evolve into dense objects in just 100 to 1000 revolutions around the Sun, and in [51, 82] the planetesimal formation rate was shown to be higher than the one obtained previously in [47]. In addition, it was found in [83] that the concentration of particles relative to the mean value increased by 3 to 4 orders of magnitude in just 15 to 30 revolutions around the Sun. According to the model in [57], the clusters that form planetesimals with more than 1 km in a radius contracted in no more than 300 years. In the computations in [84], satellite systems that formed from clusters in 100 years (i.e., in 0.6 orbital periods at a distance of 30 AU) were usually made of two or more large objects and hundreds of smaller bodies. The authors of [85] believe that, at a distance of 40 AU from the Sun, planetesimals greater than 100 km in radius formed during collapse in 25 years, the process being much slower for smaller clumps. We note that all these assessments were underestimated, because they did not take the angular momentum of the collapsing cluster into account. The dependence of the contraction time of a rarefied cluster on its angular momentum was studied by Safronov [4] and Vityazev et al. [79].

It is difficult to judge what planetesimal contraction time is closest to reality. This time depends on the conditions in a given part of the disk and on the specific parameters of the cluster under consideration. The shortest contraction times correspond to the model of solids that merge in any collision, without taking the influence of gas and angular momentum into account (this is essentially the time of free fall into the center of the cluster). These factors, as well as various interaction options during collisions of objects that make up the cluster and the dependence of their size and composition on the distance from the Sun, significantly affect the estimated cluster contraction time. We note that this time can be longer than the time of their existence, and while some clumps are just being formed in a particular zone of the disk, others are already contracted into planetesimals.

In our models of planetary accretion and body migration discussed in Sections 3–9, we study the evolution stages when the gas in the disk is exhausted and planetesimals have already formed. Therefore, the exact clump formation time and the nature of planetesimal formation (via contraction of clumps or via pebble accretion; see below) were not needed in these calculations. An exception is Section 3.4, where the contraction times of clumps were estimated in studying the formation of embryos in the Earth–Moon system and satellite systems of TNOs. It follows from the estimates presented in Section 3.4 that the time before the collision of two clumps giving rise to a parent clump for Earth–Moon embryos was most likely short, $\sim 100$ years. As regards TNOs, explaining the fraction of satellite systems formed in them requires a lifetime of clumps of the order of $10^3$ revolutions around the Sun. In other words, the lifetime of clumps in this zone of the Solar System does not exceed several hundred thousand years, and we therefore believe that the age of clumps equal to several million years in the scenario by Myasnikov and Titarenko [80, 81], which we mentioned above, is unlikely.

Based on an analysis of the properties of satellite systems, we can assume that only retrograde satellite orbits are quasistable up to the edge of the Hill sphere, while prograde orbits begin to lose stability under the influence of perturbations at about half of that sphere radius. This can be taken into account in the case of particle motion at the boundary of the Hill sphere at a low gas density. In our studies, we consider collisions of two clumps of the order of the Hill sphere in size with a low gas concentration, where the dominant role is played by the behavior of the major part of the mass. In studying the contraction of clumps, their angular momenta are taken into account in calculations by a number of authors; in particular, in Section 3.4, we discuss the effect of the angular momentum of an already formed cluster on the formation of satellite systems.

In recent years, several authors [86–91] have studied the growth of planet and planetesimal embryos via pebble accretion. Such accretion occurred in the gas and terminated when the planet embryo reached a certain mass, the so-called pebble isolation mass. Such accretion could play a much greater role in the feeding zone of the giant planets than in that of the terrestrial planets. According to [88], this can explain the dichotomy of planets at different distances from the Sun. As noted in [90], the accretion of pebbles was faster than the subsequent accretion of planetesimals; the ratio of the total mass of pebbles falling on an embryo at a distance of 5 AU from the Sun to the total mass of pebbles passing through the vicinity of the embryo was about 20%, and a large amount of matter remained in the swarm of planetesimals after the formation of giant planets [87]. The planetary embryos could migrate toward the Sun due to interaction with the gaseous disk [88, 91], but the embryo heating during their accretion prevented such migration [92].

The discussion of the results of simulations of migration processes indicates that various possible mechanisms and formation times of planetesimals could be fundamental for the formation of a circumstellar planetary system at a later stage. A special case is the Solar System. Obviously, its initial configuration was very different from the present-day one, which formed as a result of a long-term dynamic interaction of the multiple various-size initial bodies via the oligarchic growth of large planet embryos, which absorbed smaller bodies at the final stage of the formation of a diverse planetary architecture, as is observed in exoplanet systems.

This also relates to the problem of the formation of eccentricities and inclinations of planetary rotation axes. If the angular momentum of a planet relative to its center of mass is perpendicular to its orbit plane before the collision with an impactor and the angular momentum increment acquired in the collision is perpendicular to the angular momentum itself, then the ratio of the mass of the impactor to the mass of the planet is [93]

$$\frac{m_\mathrm{I}}{m_\mathrm{pl}} \approx \frac{2.5 r_\mathrm{pl}\, \chi \tan I}{\alpha\, v_\mathrm{par}\, T_\mathrm{pl}(1+\tan^2 I)^{1/2}}\,, \tag{1}$$

where $r_\mathrm{pl}$ is the radius of the planet, $T_\mathrm{pl}$ is the planet axial rotation period, $\alpha v_\mathrm{par}$ is the tangential component of the collision velocity, $v_{\mathrm{par}}$ is the surface parabolic velocity of the planet, and $\tan I$ is the tangent of the angle $I$ between the planet axis of rotation after the collision and the perpendicular to the planet orbit plane. The moment of inertia of the planet is $0.4 \chi m_{\mathrm{pl}} r_{\mathrm{pl}}^2$. For $\chi = \alpha = 1$, $T_{\mathrm{pl}} = 24$ h, and $I = 23.44°$, relation (1) gives $m_{\mathrm{I}}/m_{\mathrm{pl}} \approx 0.0065$. Hence, the present-day inclination of Earth's rotation axis could be caused by its collision with an impactor with a mass of about $0.01 m_{\mathrm{E}}$, where $m_{\mathrm{E}}$ is Earth's mass. Presumably, the collision with an impactor of a much larger mass led to the inclination of the rotation axis of Uranus, which lies almost in the orbital plane (deviating from it by of 2.23°), and also to the 'topping over' accompanied by deceleration and reversal of the sense of proper rotation of Venus [94, 95]. The orbit parameters and the tilts of rotation axes are likely to be even more diverse in exoplanet systems.

### 2.4 Timeline of planetary formation and estimates of the age of the Solar System

Scenarios of the growth of initial solids and the formation of planetesimals are closely related to the timeline of planetary formation and to estimates of the age of the Solar System. The terrestrial planets were formed by aggregation of planetesimals, as were the embryos (cores) of Jupiter, Saturn, Uranus, and Neptune, which later acquired their gas and ice shells, while the formation of the terrestrial planets was completed even later. The most accurate determination of the Solar System's age is provided by the dating of refractory calcium-aluminum inclusions (CAIs) of micrometer to millimeter size formed during the crystallization of meteorites. They are likely to belong to ancient solid matter that was part of the primary composition of the protosolar nebula, which allows determining the absolute age of the Solar System from the first condensed dust particles to the present day. The dating of CAIs in primitive meteorites by different groups of authors using corrected U–Pb and Pb–Pb analyses gave close values of $4567.22 \pm 0.21$ Myr and $4568.67 \pm 0.17$ Myr (the latter value is believed to be more accurate [96–100]). At the same time, the age of chondrules is in the range of $4567.32 \pm 0.42$ to $4564.71 \pm 0.30$ Myr, which indicates that chondrules were formed almost simultaneously with CAIs and this process took $\sim 3$ Myr. This period of time is close to the time of the existence of the protoplanetary accretion disk, whose secular evolution is obviously directly related to this process.

At the same time, the absolute age of iron meteorites has been determined as $4567.5 \pm 0.5$ Myr. Therefore, given the error range, the time of the origin of the Solar System is determined up to $\sim 1$ Myr, or with an accuracy of 0.002%. The oldest anorthositic rocks of the Moon and terrestrial zircons are only slightly younger: they are estimated to be $\sim 4.4$ billion years old. The absolute age of stony meteorites is $4564.91 \pm 2.58$ Myr. The difference $\Delta = 3.64 \pm 1.52$ Myr can be regarded as an estimate of the total time of formation and differentiation in the course of thermal evolution of the parent bodies of these ancient meteorites. Based on the totality of available data, it can be assumed that the Solar System is $4567.3 \pm 0.1$ billion years old [101].

On the whole, the time scale outlined above is consistent with computer simulation results. They give grounds to believe that, while the accretion of disk matter on the proto-Sun was completed in 1–2.5 Myr after the birth of the protoplanetary system, the dust subdisk, consisting of approximately centimeter-size particles, formed much earlier, in 0.1 to 1.0 Myr at the radial distance $R \sim 1$ AU, where gravitational instability developed after the critical density was reached. Obviously, the subsequent $\sim 1$–2 Myr sufficed for the formation and thermal evolution of the first solids.

## 3. Migration of planetesimals during the formation of the terrestrial planets

### 3.1 Modeling an isolated aggregation of the terrestrial planets

As already noted, many authors (see, e.g., [4, 34, 79, 102, 103]) assumed that the terrestrial planets and the cores of the giant planets were formed from a swarm of solid bodies — planetesimals moving around the Sun along initially nearly circular orbits. The process of the formation of the terrestrial planets was first studied analytically [104–131]. In particular, the time of the formation of the Earth was estimated at about 100 Myr. At present, a large number of papers have been published on numerical simulations of the formation of the terrestrial planets from planetesimals [132–159], mostly dealing with the model of the evolution of disks of gravitating bodies that merge in collisions in the feeding zone of terrestrial planets. The actual planet formation process was very complex and depended on many factors, but studies relying on relatively simple models allow capturing a number of important features of the process. In [35, 36, 160–163], numerical simulations were used to study the formation of protoplanets by the merger of highly rarefied gas–dust clumps moving in almost circular orbits. The clumps were assumed to merge into giant rarefied protoplanets of the masses of the modern planets even before they were compressed to the density of solid bodies.

In 1978, the first studies of the evolution of solid-body rings appeared, where gravitational interactions of the disk bodies were modeled in a planar model and two bodies were assumed to merge when the distance between their centers of mass became equal to the sum of their radii (rather than the radius of some large conventional sphere) [164, 165]. A spatial model of the evolution of such disks was first considered in [102]. In [102, 137], the number of initial bodies was 100, and the method of spheres was used to describe the gravitational influence of the bodies. The evolution of disks containing up to 1000 gravitating bodies each in the feeding zone of the terrestrial planets, which merged during collisions, was considered by Ipatov [103, 141–145]. Their mutual gravitational influence was taken into account within the method of the spheres of action, when bodies are assumed to move around the Sun along unperturbed Keplerian orbits outside the sphere, whereas their relative motion inside the sphere is determined by a two-body problem. The initial distances from the bodies to the Sun ranged from 0.36–0.40 AU to 1.2 AU, and their total mass was $1.87 m_{\mathrm{E}}$, where $m_{\mathrm{E}}$ is Earth's mass. Calculations of the evolution of planar disks showed that, in the case of almost circular initial orbits, the number of formed planets is greater than four, and the actual number of terrestrial planets is obtained only at the initial eccentricities $e_0 = 0.35$. The actual number of planets can also be obtained in the model of spatial disk evolution; in one of the computations, four planets with masses greater than $0.046 m_{\mathrm{E}}$ were formed. The actual number of planetesimals was much more than 1000, and only part of the matter comprising two colliding bodies formed a new body. The computing power of 20th century computers did not allow considering more complex models, however; accounting for the fragmentation of colliding bodies could somewhat increase the planetary formation time.

In computing the evolution of spatial disks, the initial eccentricities of the orbits of the bodies were taken to be $e_0 = 0.02$, and it was shown that such eccentricities are quite rapidly achieved when taking the mutual gravitational influence of the bodies into account at distances greater than the radii of their spheres of action. In turn, the average eccentricity $e_{av}$ of the orbits of bodies in the course of evolution exceeded 0.2, and in a number of versions was greater than 0.4 at some instants. For example, in a version with $e_0 = 0.02$ and 960 initial bodies, $e_{av}$ was respectively equal to 0.09, 0.20, and 0.35 for 500, 250, and 100 bodies in the disk. Large average orbital eccentricities were observed for bodies located along the disk edges, with major semiaxes $a < 0.4$ AU and $a > 1.2$ AU, and by end of evolution, the orbits of some planets with masses of the order of the masses of Mercury and Mars acquired eccentricities close to the actual eccentricities of these planets. We note that the increase in the eccentricities of Mercury and Mars (and the inclination of the orbit of Mercury) could be due to not only the influence of large bodies from the feeding zone of the terrestrial planets but also the gravitational influence of bodies that entered that feeding zone from the feeding zones of the giant planets. At the same time, these bodies themselves could avoid collisions with bodies in this feeding zone and only perturb their orbits gravitationally. The high abundance of iron in the core of Mercury is usually explained by the loss of most of the silicate shell mass in high-speed impacts. At the same time, part of the planetesimals in the vicinity of Mercury's orbit, which passed relatively close to the Sun before their collisions with Mercury's embryo, could have lost part of their silicate composition during such passages and thereby affect the high content of iron in Mercury's core.

The masses of the embryos of unformed terrestrial planets could exceed $0.05 m_E$, which allows explaining both the tilt of the rotation axis and Earth's axial rotation period. According to estimates, the time of formation of 80% of the mass of the largest stony planet (analogous to Earth) did not exceed 10 Myr, while the total time of evolution of the disks of gravitating bodies was about 100 Myr. In computations by the method of spheres of action, times of the order of 1–10 Myr were obtained for the formation of the major part of planet mass by considering a 'deterministic' method for selecting pairs of approaching bodies: a pair of bodies with the minimum approach time was selected in modeling [103, 166]. Meanwhile, the use of a 'probabilistic' method for selecting pairs of approaching bodies (proportionally to the probability of their approach) yielded formation times of most of the planetary mass almost an order of magnitude longer. The time it took for the number of bodies in the disk to decrease from $N_0$ to $N$ was usually about half the time it took for the number of bodies to decrease from $N_0$ to $N/2$, and most of the evolution of the disks was taken by the last stages of planetary formation. It was therefore concluded that the total disk evolution time for $N_0 = 10^{12}$ is approximately the same as for $N_0 = 10^3$, but taking collisional fragmentation of bodies into account can lead to a severalfold increase in the time needed for the major part of a planet's mass to accumulate [103, 111].

As shown by more recent results of numerical integration of the equations of motion with the mutual gravitational influence of bodies taken into account more thoroughly [133–136, 139, 140, 147, 149, 151, 153–155], the deterministic approach reflects the actual evolution of the disks of bodies and the planetary formation times quite satisfactorily (and better than the probabilistic approach). In the first such computations [136], 56 planetary embryos were considered and the computations required about three years of processor time, but the number of bodies simulated recently lies in the thousands (as, for instance, in simulations with 6000 planetesimals in [149]), which evidently improves the statistics. The main conclusions about the formation of the terrestrial planets drawn from the calculations by the method of spheres of action and by numerical integration of the equations of motion are approximately the same.

### 3.2 Modeling the aggregation of the terrestrial planets with the influence of the giant planets taken into account

The formation and migration of the giant planets are closely related to the aggregation of the terrestrial planets. Planetesimals from the feeding zone of the giant planets, which were acquiring orbits with small perihelion distances during the Solar System's evolution, perturbed the orbits of planetesimals and planet embryos in the feeding zone of the terrestrial planets, as well as bodies of the asteroid belt, and often collided with them. Changes in Jupiter's and Saturn's orbits caused changes in the positions of resonances and contributed to the cleaning of the asteroid belt zone, from which some bodies could penetrate into the feeding zone of the terrestrial planets. Therefore, the influence of the forming giant planets and their feeding zone bodies must be taken into account when studying the accretion of the terrestrial planets.

Various scenarios of such influence have been considered. In [153–155], the impact of Jupiter on the formation of the terrestrial planets was simulated for various values of its orbit and mass. In these computations, the terrestrial planets reached half of their final masses in the first 10 to 20 Myr, although separate bodies continued falling on them after 100 Myr. In [154, 155], the initial disk with 1000 to 2000 planetesimals was considered, a number 5 to 10 times greater than in previous papers where the mutual gravitational influence of bodies was taken into account by numerically integrating the equations of motion. The initial disk with a mass of $9.9 m_E$ extended up to 5 AU. Over a billion years, the asteroid belt was cleared by more than 99% due to resonances of planetesimals with Jupiter determined by their mutual gravitational influence and the influence of embryos. It was noted in [140] that, as the gas dissipated, the secular resonances ($v_5$, $v_6$, $v_{15}$, and $v_{16}$) with Jupiter and Saturn moved inward, affecting the planetesimals. After 3 Myr, the gaseous disk decreased in mass by a factor of 20 and did not produce a dynamical effect on the migration of planetesimals.

A number of authors studying the formation of the terrestrial planets considered the Grand Tack model (see, e.g., [167–170]). In this model of the early dynamical rearrangement of the Solar System, interaction with the gas in the disk first made Jupiter migrate closer to the Sun, up to 1.5 AU; then, after the formation of a massive Saturn and the scattering of gas, Jupiter, together with Saturn, started moving away from the Sun, staying in a 2:3 resonance with Saturn. As a result of this migration, Jupiter 'cleared' the asteroid belt, reduced the amount of material in Mars's feeding zone, and facilitated the delivery of water to the forming terrestrial planets. It was assumed in [170] that Jupiter and Saturn acquired rather large masses of gas over 600 thousand years and respectively migrated from 3.5 and 4.4 AU to 1.5 and 2 AU during the first 100 thousand years;

after the mass of Saturn increased from $10m_E$ to its present-day mass, Jupiter and Saturn migrated to respective distances of up to 5.25 and 7 AU from the Sun in 500 thousand years. In Section 4, we discuss models treating the subsequent stages of the agglomeration of the giant planets and the migration of Uranus's and Neptune's embryos, including the Nice model, named after the place of its creation, the French observatory in Nice. In the Nice model, the cause of abrupt changes in the orbits of these embryos is assumed to be the 1:2 orbital resonance established between Jupiter and Saturn.

The mechanism of giant planet migration explains a number of events in the early history of the Solar System, including the presence of numerous TNOs in resonances with Neptune, the Kuiper belt, and Oort cloud formation, and phenomena such as the hypothetical late heavy bombardment of the inner Solar System, although its proposed mechanism is not universally accepted, despite being used to estimate the role played by an exogenous source of volatiles in the evolution of the terrestrial planets. The delivery of water to the Earth by small bodies is assumed to have occurred mainly after these planets acquired 60 to 80% of their final mass. According to other estimates [171], volatile components in Earth's zone could be accumulated by parent planetesimal bodies only 1 Myr after the formation of the protoplanetary circumsolar disk, when its temperature dropped to below 700 K. As we discuss in what follows, in various models (for example, the one in [169]), the growth time of an Earth analogue to $0.5m_E$ was in the range from several to 20 Myr.

Studies within the Nice model included the evolution of the orbits of asteroids during an abrupt change in the orbit of Jupiter, leading to a sharp change in the positions of resonances [172]. The probability of collisions of asteroids with the Moon was obtained equal to $4 \times 10^{-5}$, and 20 times higher for the Earth. According to [173, 174], the Nice model explains the formation of Mars and the asteroid belt well if the above instability occurred within 1 to 10 Myr after the gaseous disk dissipation.

In contrast to the studies discussed in Section 3.1 of the evolution of disks of bodies that merge in collisions, another model was used in [146] to study the formation of the terrestrial planets. Migration of planetesimals was computed within the feeding zone of the terrestrial planets, divided into seven regions, depending on distance from the Sun. The gravitational influence of all planets, including the giants, was taken into account, while the planetesimals and the planets themselves were regarded as point masses and their collisions were not taken into account directly. In a number of versions of the model, embryos with masses from 0.1 to 0.3 of modern planet masses were considered instead of the terrestrial planets. The arrays of orbital parameters of migrating planetesimals obtained in the computations with a step of 500 years were used to calculate the probabilities of their collisions with the planets, their embryos, and the Moon. This approach allowed more accurately calculating the probabilities of collisions between planetesimals and planetary embryos for a number of evolutionary stages, especially if these probabilities are small. Later, we carried out computations similar to those in [146] for a model where the planetesimals that have collided with a planet are excluded from further computations.

When studying the composition of the planet embryos from planetesimals initially located at various distances from the Sun, narrower zones of planetesimal origin were considered than were in previous studies, and not only the final composition of the planets but also the change in the composition of the embryos over time was studied. The conclusion drawn from the computations was that terrestrial planet embryos with masses of the order of or less than one tenth of the modern planet masses were mainly accumulating planetesimals from the vicinity of their orbits. The inner layers of a terrestrial planet were formed mainly from matter from the vicinity of the planet orbit. When planetesimals fell out of Jupiter's and Saturn's feeding zone onto terrestrial planet embryos, these embryos had not yet acquired the masses of modern planets, and matter (including water and volatiles) from this zone could have fallen into the inner layers of the terrestrial planets and affected their composition. With the masses of Earth's and Venus' embryos of the order of a third of their modern masses, the probabilities of fallouts of planetesimals formed at a distance from 0.7 to 0.9 AU from the Sun onto these embryos differed by not more than a factor of two in the considered time interval $T > 2$ Myr.

Based on the considered model, it was also found that the total masses of planetesimals that migrated from each zone in the region from 0.7 to 1.5 AU from the Sun and collided with almost-formed Earth and Venus differed by not more than a factor of two. The outer layers of Earth and Venus could have accumulated the same material from different parts of the feeding zone of the terrestrial planets. At the final formation stages, the planetesimals initially located at 1.1 to 2.0 AU from the Sun could have become part of Earth and Mars in a ratio not much different from that of the masses of these planets.

In [146], the fraction of planetesimals that fell on the Sun could exceed 10% for the initial planetesimal distances from the Sun in the range from 0.3 to 0.5 AU and from 1.1 to 2.0 AU. In the versions where planetesimals that collided with planets were excluded from computations, the evolution time of planetesimal disks was typically equal to several hundred Myr. But, in some versions of computations with small initial eccentricities, individual planetesimals were moving in 1:1 resonances with Earth or Venus even after a billion years or more. The time interval considered in these versions of computations was longer than the one in [146], and the proportion of bodies that collided with the Sun, given the present-day planetary masses, was mainly in the range of 0.24–0.32 for the initial semimajor axes of planetesimals $a_0 < 1.1$ AU, and reached 2/3 for $1.5 \leqslant a_0 \leqslant 2$ AU. In most cases, for planet masses no greater than half their modern masses and for a time interval not exceeding 10 Myr, no planetesimals collided with the Sun or were ejected into hyperbolic orbits.

As in earlier computations [103, 142–145], the proportion of planetesimals ejected from the feeding zone of terrestrial planets into hyperbolic orbits did not exceed 10%. At the same time, the probability of a collision with Jupiter for a planetesimal initially located in the feeding zone of the terrestrial planets was no more than a few percent of the probability of its collision with Earth, and the probability of a collision with Saturn was even lower by an order of magnitude.

The above model estimates of the formation of terrestrial planet embryos were based on a model that takes the joint gravitational influence of giant planets and terrestrial planet embryos into account. Accounting for the mutual gravitational influence of planetesimals, including those that come from the feeding zones of the giant planets, leads to an increase in the mixing of matter in the feeding zone of the terrestrial planets and an increase in the probability of collisions of planetesimals with the Sun and their ejection into hyperbolic orbits. For the mass ratio of Earth and Moon embryos equal to 81, similar to the modern one, the ratio of the probabilities of planetesimals falling onto Earth and Moon embryos in the considered versions did not exceed 54, and it was maximum for embryo masses of approximately one third the modern masses of these bodies.

In recent years, the formation of the terrestrial planets has mainly been studied based on the Nice and Grand Tack models mentioned above. As we have seen, the first model is based on the hypothesis of an abrupt change in giant planet orbits when Jupiter and Saturn establish a resonance, and Jupiter's migration to the orbit of Mars and back is assumed in the second model. It was noted in [146], however, that the peculiarities of the formation of the terrestrial planets and the clearing of the asteroid belt can be explained without using these models, based solely on a relatively smooth decrease in Jupiter's semi-major axis and a shift in the positions of resonances due to ejection of planetesimals into hyperbolic orbits by Jupiter. In such a model, the embryo formation of Uranus and Neptune near Saturn's orbit is assumed and the migration of these embryos to the modern orbits of Uranus and Neptune due to interaction with planetesimals is considered [103, 145, 175, 176].

### 3.3 Formation times of the terrestrial planets

When discussing model approaches in the preceding sections, we already touched upon estimates of the time scale of agglomeration of the terrestrial planets. We now consider this problem in more detail, invoking the results of studying the isotope composition of meteorites.

From analyses of the lead isotope ratio $^{207}$Pb/$^{206}$Pb in zircon crystals contained in a substance from the Martian meteorite NWA 7034 [177], it was concluded that the formation of the core and the crystallization of a magma ocean on Mars were completed no later than 20 Myr after the beginning of the formation of the Solar System, measured from the formation of CAIs. This estimate is consistent with the hafnium–tungsten scale $^{182}$Hf–$^{182}$W, indicating the age within 10 Myr [178]. The obtained estimate of the Hf/W ∼ 4 isotope ratio for the Martian mantle with a ∼ 25% uncertainty corresponds to the Martian core formation time in the range from 0 to 10 Myr [179]. Thermal models also indicate that the solidification of Mars was completed within ∼ 10 Myr [180], during which Mars grew to approximately its present size. It was assumed in [181] that agglomeration of Mars was completed within approximately 5 Myr. According to [146], Mars grew more slowly than Earth and Venus, and individual planetesimals could remain in its feeding zone even after 50 Myr. It can therefore be assumed that, during clump contraction, a rather large Martian embryo with a mass of at least $0.02m_{\rm E}$ was initially formed, and planetesimals from Jupiter's and Saturn's feeding zones contributed to a more rapid removal of planetesimals from Mars' feeding zone. A similar scenario has been proposed for the formation of the Mercury embryo with the same initial mass.

According to the model in [149], in a disk with a mass of ∼ $7m_{\rm E}$ in the range of 0.2–3.8 AU from the Sun, the average (over several calculation schemes) mass of a Mercury analogue was about $0.2m_{\rm E}$, greatly exceeding the mass of Mercury. The main contributors to the mass of a Mercury analogue were bodies from the zone at 0.2 to 1.5 AU over 10 Myr, later supplemented by bodies from the zone at up to 3 AU from the Sun. The orbits of these analogues had semimajor axes close to 0.27–0.34 AU, and their eccentricities and inclinations were small. According to [139], considering the evolution of the disk at a distance from 0.7 to 1.0 AU from the Sun shows that the analogues of Earth and Mars accumulated most of their mass in 10 Myr. Individual planetesimals could fall on the forming terrestrial planets until 100 Myr [118], which is confirmed by numerical calculations [102, 136, 142–145].

Of particular interest are Earth and Venus. Based on the model of accumulation of bodies in their feeding zones in any sort of collisions, the masses of their embryos could double in 1 Myr, respectively starting from $0.1m_{\rm E}$ and $0.08m_{\rm E}$. We note that the bodies from Jupiter's feeding zone start penetrating into the feeding zone of the terrestrial planets in the same period. Earth and Venus could have acquired more than half their masses in approximately 5 Myr. The 3 to 5 Myr estimate for the time scale is supported by data from studies of the isotopic composition of Earth's atmosphere [182]. During that time, most of the planetesimals initially located at a distance of 0.7 to 1.1 AU from the Sun fell on the growing Earth and Venus [146]. Obviously, taking the release of matter during collisions into account would lead to an increase in the accretion time.

As already noted, at the initial stages of the Solar System's evolution, the residual gas in the disk played an important role. When studying the formation of bodies in the zone from 0.5 to 4 AU in [140], it was assumed that the surface density of the gas decreased exponentially with time, $\Sigma_{\rm gas}(r, t) = \Sigma_{\rm gas,0}(1 {\rm ~AU}, 0) (r/1 {\rm ~AU})^{-1} \exp{(-t/\tau)}$. It was assumed in these calculations that $\tau = 1$ Myr and $\Sigma_{\rm gas,0} = 2000$ g cm$^{-2}$. After 4.6 Myr, only 1% of the gas remained. The gas could possibly dissipate in a time ∼ 10 Myr [183], and therefore the agglomeration of planetesimals occurred in the presence of the gas component, which reduced the eccentricities of the planetesimals; the formation time of the terrestrial planets did not exceed 10 Myr either.

The evolution of narrow planetesimal rings 0.02 AU and 0.092 AU wide at a distance of about 1 AU from the Sun, with and without taking the gas drag into account for bodies moving in the rings, was studied in the models in [184, 185]. It was shown that the distance between the orbits of the emerging planetary embryos was about $(5–10)r_{\rm H}$, where $r_{\rm H}$ is the radius of the embryo's Hill sphere. The formation time of protoplanets with masses of ∼ $10^{26}$ g (∼ $0.016m_{\rm E}$) was about 0.5 Myr, and after 1 Myr the main mass of the disk was concentrated in bodies with a mass of no less than several $10^{26}$ g. In semianalytic models [134], the formation of a $0.1m_{\rm E}$ embryo at a radial distance of 1 AU and of a $10m_{\rm E}$ embryo at 5 AU, respectively, takes 0.1 and 1 Myr. Upon reaching a mass of ∼ $10m_{\rm E}$, Jupiter's embryo could relatively quickly continue increasing its mass by gas accretion.

As shown by numerical studies of the migration of planetesimals from Jupiter's and Saturn's feeding zones, a major part of their mass left this zone in several million years. This indicates the time when the planetesimals coming from the zones of Jupiter and Saturn influenced the formation of the terrestrial planets. We emphasize that, according to the calculations performed in the framework of the model that takes the interaction of all Solar System planets into account [186], the maximum evolution times of the orbits of some planetesimals that started in the zone of Jupiter and Saturn can be much longer than in the absence of Uranus and Neptune (50 and 4 Myr, respectively). Even in that case, however, the main contribution to collisions with the terrestrial planet embryos was made by planetesimals from Jupiter's and Saturn's feeding zones in the first million years after the formation of a significant mass of Jupiter. This time is estimated as 1 to 2 Myr from the origin of the Solar System. Meanwhile, individual planetesimals from Uranus's and Neptune's feeding zones fell on the Earth even after hundreds of millions of years, and may even remain in the Solar System to this day. In the Grand Tack model, when planetesimals fell out of Jupiter's and Saturn's feeding zones and from the zone of the outer asteroid belt onto the terrestrial planet embryos, these embryos had not yet acquired the modern planetary masses, and material (including water and volatiles) from these regions could accumulate in the inner layers of the forming terrestrial planets and the Moon.

### 3.4 Formation of the Earth–Moon system

Historically, various models of the origin of the Moon have been proposed, but this problem is not yet decisively resolved. We discuss several of the most significant options here.

According to the theory of coaccretion (see, e.g., [187–190]), the Moon formed from a near-Earth swarm of small bodies. Their main source, according to the Schmidt–Ruskol–Safronov model, was the sticking together of protoplanetary disk particles during collisions ('free to free' and 'free to bound').

The mega-impact model proposed by a number of authors [191–199] has won considerable popularity. The Moon is assumed to have been formed in a catastrophic collision with a Mars-size body (named Theia), which caused Earth's molten silicate mantle to be ejected into low Earth orbit. A proto-Moon formed from the merged fragments of the ejection gradually receded to its present-day orbit due to tidal interaction with Earth in the course of evolution. The attractiveness of this hypothesis consists primarily in its explaining the average density of the Moon, which is equal to the density of Earth's mantle. Several modifications of the mega-impact model have been proposed. In particular, calculations in [197] showed that, when a body of a mass from $0.026m_E$ to $0.1m_E$ hits a proto-Earth that is rapidly rotating with a period of about 2.5 h, a Moon-forming disk can emerge, consisting mainly of the substance of Earth's mantle. According to [194], a head-on collision of two bodies of almost equal masses (with a mass ratio not greater than 1.5) could give rise to similar compositions of Earth and the Moon. We emphasize, however, that these models require the subsequent removal of a part of the angular momentum of the formed Earth–Moon system by means of an orbital resonance between the Sun, Earth, and the Moon.

The canonical mega-impact model encounters certain difficulties, primarily of a geochemical nature, and is currently being critically scrutinized. It is not capable of explaining the similar isotopic abundances of a number of elements on the Earth and the Moon, primarily oxygen, iron, hydrogen, silicon, magnesium, titanium, potassium, tungsten, and chromium [199–204]. It can hardly be assumed that the body that formed the Moon, even if coming from a relatively close vicinity of the forming terrestrial planets, had a composition similar to that of Earth, because, according to this model, most of the Moon's substance originates from the impactor rather than the proto-Earth. These results undermine geochemical substantiation of the mega-impact model. In addition, the mega-impact hypothesis does not offer a means of explaining the absence of isotopic shifts in lunar and terrestrial matter, because the material ejected during a giant impact should be 80–90% vapor, and the isotopic compositions of K, Mg, and Si should change noticeably when the melt evaporates [201]. The giant impact hypothesis suggests that, after the collision, a magma ocean formed on Earth's surface, but the connection of the ancient magma ocean on Earth with this event is debatable [205].

An alternative to the mega-impact model is the multi-impact model [206–212] and the model of the formation of Earth and Moon embryos from a single initial rarefied gas–dust cluster in a protoplanetary nebula, followed by the formation and contraction of two fragments [200–202, 213–217].

The multi-impact model is based on the hypothesis of multiple collisions (macroimpacts) of planetesimals with Earth's embryo. It was found in [209] that matter ejected from the Earth into prograde orbits easily joins the prograde protosatellite disk, while matter ejected from the Earth into retrograde orbits falls back to the Earth. In [212], up to $10^6$ particles were included in calculations of the impactor collision with Earth's embryo by the method of smoothed-particle hydrodynamics (SPH). In this method, the liquid is regarded as a set of discrete moving elements, particles. The impactor mass varied from 0.01 to 0.09 of the Earth embryo mass, and the collision velocity varied from 1 to 4 parabolic velocities on the embryo surface. The collision angle and the angular velocity of rotation of the embryo were also varied. The fraction of iron in the formed Moon did not exceed 10% in 75% of the cases considered.

The model of the formation of the Earth and Moon embryos from a single initial rarefied gas–dust clump in the protoplanetary nebula with the subsequent formation and contraction of two fragments [200–202, 213–217], in addition to satisfying geochemical constraints, also allows explaining the known differences in the chemical composition of Earth and the Moon, including iron deficiency, depletion of volatiles, and enrichment in refractory oxides of Al, Ca, and Ti of the Moon compared to Earth.

The bound of no more than 30 Myr for the formation of the major part of the mass of Earth and the Moon has been obtained by studying the hafnium–tungsten Hf/W isotope abundance [218, 219]. Based on the study of the neon $^{20}$Ne/$^{22}$Ne ratio, it was concluded that the presence of nebular neon requires Earth's embryo to acquire a significant mass in a few million years and to be able to capture the nebular gases that had dissolved in the ancient magma ocean [220]. In contrast to the studies cited in Section 3.3, where relatively short formation times for most of the mass of terrestrial planets were assumed based on the analysis of the $^{182}$Hf–$^{184}$W and $^{87}$Rb–$^{86}$Sr systems, Galimov [215] came to the conclusion that the formation of the Earth and Moon cores could not have begun earlier than 50 Myr from the CAI-dated origin of the Solar System. It has also been suggested that, before the Moon formed as a condensed body, it must have evolved in an environment with a higher Rb/Sr ratio. Due to its large atomic weight, rubidium cannot escape from the surface of the Moon, but can escape from the heated surface of small bodies or particles, and therefore the original lunar substance had probably remained in a dispersed state for the first 50 Myr, for example, in the form of a gas–dust clump.

The model suggested by Galimov et al. is not free of shortcomings either; we discuss it critically in [221]. In particular, it remains undetermined where the iron-depleted substance has gone from the inner part of the clump until the moment when the embryo started growing in its outer part. Under the condition of the inflow of matter from outside the Hill sphere, the model with zero relative velocities considered in [202] could hardly be realized: even inside the clump, the particle velocities could not be zero in the presence of rotation. At zero velocities, the particles would very quickly (with a free-fall time equal to 25 years according to [57]) fall onto the center of the clump before the embryos formed in its hot inner part, where the evaporation of particles alone took tens of thousands of years [202]. We therefore believe that the existence of the clump that gave rise to the Earth and the Moon within 50 Myr is unlikely, which is corroborated by a review of the studies of clump lifetimes in Section 2.3. As noted in Section 3.3, the time of the essential formation of the Earth probably did not exceed 5 Myr.

In contrast to the model of Galimov et al., where the mass of the initial clump was equal to the total Earth plus Moon mass, the embryos of Earth and the Moon in Ipatov's model [221] were formed from a common rarefied clump with a mass greater than $0.01m_E$ and an angular momentum sufficient for the formation of the Moon embryo. The angular momentum of the clump necessary for the formation of Earth–Moon system embryos was acquired in the collision of two clumps. The growth of the Earth and Moon embryos was considered within the multi-impact model. Most of the matter that became part of the Moon embryo, mainly moving in the vicinity of the Earth, was ejected from the Earth during its numerous collisions with planetesimals and smaller bodies. Some minor bodies also have satellites [222]; a mechanism analogous to the model in [221] was proposed for the model of the formation of trans-Neptunian satellite systems [223–225].

Let us consider this scenario in more detail. It is based on the collision of two spherical clumps with radii $r_1$ and $r_2$ moving before the collision in the same plane in circular heliocentric orbits with the difference between their semi-major axes $a$ equal to $\Theta(r_1+r_2)$, without taking their mutual gravitational influence into account. In this case, the tangential component of the collision velocity is $v_\tau = v_c(r_1+r_2)a^{-1}k_\Theta$. For $(r_1+r_2)/a \ll \Theta$, we have $k_\Theta \approx 1-1.5\Theta^2$. It follows that $k_\Theta$ can take values from $-0.5$ to $1$. The average value of $|k_\Theta|$ is 0.6. With this $k_\Theta$, the collision velocity of the clumps is

$$v_{col} = v_c(r_1+r_2)a^{-1}(1-0.75\Theta^2)^{1/2}, \quad (2)$$

and the tangential component of the collision velocity is

$$v_\tau = v_c(r_1+r_2)a^{-1}(1-1.5\Theta^2) = v_c(r_1+r_2)a^{-1}k_\Theta, \quad (3)$$

where $v_c = (GM_S/a)^{1/2}$ is the heliocentric velocity of the clump, $G$ is the gravitational constant, and $M_S$ is the solar mass [223].

Given formula (3), it follows [223] that, before the collision, the angular momentum of two colliding clumps (with radii $r_1$ and $r_2$ and masses $m_1$ and $m_2$) moving in circular heliocentric orbits with semi-major axes close to $a$ is equal to

$$K_s = k_\Theta(GM_S)^{1/2}(r_1+r_2)^2 m_1 m_2(m_1+m_2)^{-1}a^{-3/2}. \quad (4)$$

The values of $K_s$ and $k_\Theta$ are positive for $0 < \Theta < 0.8165$ and negative for $0.8165 < \Theta < 1$. In the case of a collision of two identical clumps whose radii are equal to $k_H r_H$ (where $r_H$ is the Hill sphere radius for the clump with a mass $m_1 = m_2$), it follows from formula (4) that

$$K_{s2} = K_s \approx 0.96 k_\Theta k_H^2 a^{1/2} m_1^{5/3} G^{1/2} M_S^{-1/6}. \quad (5)$$

Using formula (5), we can see that $K_{s2}$ is equal to the present-day angular momentum $K_{\Sigma EM}$ of the Earth–Moon system for $k_\Theta = 1$ and $2m_1 \approx 0.096 m_E$. Thus, the angular momentum of the Earth–Moon system could be acquired in a collision of two clusters (moving in circular heliocentric orbits before the collision) with a total mass not less than the mass of Mars. As shown in [221], the initial mass of the rarefied clump that gave rise to the Earth and Moon embryos could be relatively small ($0.01m_E$ or even less) if the increase in the angular momentum of the embryos due to the growth of their masses is taken into account. For nonzero eccentricities of the heliocentric orbits of the clumps, the angular momentum acquired in their collision can be greater than for the circular heliocentric orbits considered above. Part of the mass and the angular momentum is lost in the collision (especially in a tangential collision) and the contraction of the cluster. Therefore, the mass and angular momentum of the colliding clusters could be greater than those of the resulting parent clump and of the satellite system formed during its contraction.

Taking into account that $K_s = J_s \omega_c$, we discover from formula (4) that the angular velocity of the clump formed in the collision of two clumps is

$$\omega_c = 2.5 k_\Theta \chi^{-1}(r_1+r_2)^2 r^{-2} m_1 m_2(m_1+m_2)^{-2}\Omega, \quad (6)$$

where $\Omega = (GM_S/a^3)^{1/2}$ is the heliocentric angular velocity of the clump. The moment of inertia of the resulting clump of radius $r$ and mass $m$ is $J_s = 0.4\chi mr^2$, where $\chi$ characterizes the distribution of matter inside the clump ($\chi = 1$ for a homogeneous spherical clump, considered in [84]). For $r_1 = r_2$, $r^3 = 2r_1^3$, $m_1 = m_2 = m/2$, and $\chi = 1$, we have $\omega_c = 1.25 \times 2^{1/3} k_\Theta \Omega \approx 1.575 k_\Theta \Omega$.

In calculations [84] of the contraction of clumps (of mass $m$ and radius $r$) residing in the trans-Neptunian region, their initial angular velocities were taken equal to $\omega_0 = k_\omega \Omega_0$, where $\Omega_0 = (Gm/r^3)^{1/2}$ is the circular (orbital) velocity on the surface of the cluster. We note that $\Omega_0/\Omega = 3^{1/2}(r_H/r)^{3/2} \approx 1.73(r_H/r)^{3/2}$; if $r \ll r_H$, then $\Omega \ll \Omega_0$. In the case of Hill spheres, assuming the angular velocity $\omega_c \approx 1.575 k_\Theta \Omega$ of the clump formed by the collision of two identical clumps to be equal to $\omega_0$, we have $k_\omega \approx 0.909 k_\Theta/\chi$. This relation shows that collisions of clumps with $k_\Theta = \chi = 1$ can yield values of $\omega_c = \omega_0$ corresponding to $k_\omega$ values up to 0.909. In [84], binary or triple systems were obtained only for $k_\omega$ equal to 0.5 or 0.75. We can therefore conclude that the initial angular velocities of clumps at which binary systems were formed could be gained in their collisions. According to [4], the initial angular velocity of a rarefied clump with respect to its center of mass is $0.2\Omega$ for a spherical cluster and $0.25\Omega$ for a flat disk. The initial angular velocity is always positive and can be almost an order of magnitude less than the angular velocity acquired in the collision of clumps. Because $\Omega_0/\Omega \approx 1.73(r_H/r)^{3/2}$, it follows that for $r = r_H$ we have $\Omega \approx 0.58\Omega_0$ and the initial angular the rotation velocity of a rarefied spherical clump with respect to its center of mass is $0.2\Omega \approx 0.12\Omega_0$. If $r \ll r_H$, then $\Omega \ll \Omega_0$. It follows from the above estimates that the angular velocity and angular momentum of the clump formed directly from a protoplanetary disk were insufficient for the formation of a satellite system.

We note that two clumps whose collision resulted in forming the clump whose contraction gave rise to the Earth and Moon embryos could have moved around the Sun in different planes before the collision, and therefore the orbital plane of the Moon embryo could differ from the ecliptic plane and the existing inclination of 5.1° was formed. Ipatov [224] showed that the characteristic paths of the constituents of the colliding clumps before the collision with objects of another clump are shorter than the size of the clumps, which indicates the possibility of clump merger during their collision. In the same study, estimates were obtained for the angular momentum $K_s$ of a clump of mass $m_f$ that grew by accumulating smaller objects. For radius $r$ of the growing cluster equal to $k_H r_H$ (where $k_H$ is a constant and $r_H$ is the Hill radius of the growing cluster) and $|v_\tau| = 0.6 v_c r a^{-1}$, we have

$$K_s \approx 0.173 \, k_H^2 \, G^{1/2} a^{1/2} m_f^{5/3} M_S^{-1/6} \Delta K, \qquad (7)$$

where $M_S$ is the solar mass and $\Delta K = K^+ - K^-$ is the difference between the positive $K^+$ and negative $K^-$ angular momentum increments for the clump when small celestial objects fall on it ($K^+ + K^- = 1$). Formula (7) was obtained by integrating the angular momentum increment with respect to the mass $m$ from 0 to $m_f$. It was taken into account that the angular momentum increment is equal to $dK_s = r v_\tau \, dm$, with $dm = 4\pi \rho r^2 \, dr$ and $m = 4\pi \rho r^3/3$ (density $\rho$ of the growing clump was assumed constant). When considering the growth of the clump mass from $m_0$ to $m_f$ in formula (7), $m_f^{5/3}$ is replaced with $(m_f - m_0)^{5/3}$.

It is interesting to compare the angular velocity of the clump acquired in the course of accumulation of smaller objects with the angular velocity $\omega_0$ required for the formation of a satellite system under clump contraction. Comparing $K_s = J_s \omega_0$ (where $\omega_0 = k_\omega \Omega_0$ and $J_s = 0.4 \chi m r^2$) with the $K_s$ calculated by formula (7), we obtain $\Delta K \approx 0.8 \chi k_\omega$ (for any $r$ and $m$). It follows from this relation that, for $\chi = 1$, $\Delta K$ is approximately equal to 0.4, 0.5, and 0.6 for the respective values of $k_\omega$ given by 0.5, 0.6, and 0.75. The values of $\Delta K$ are usually smaller for colliding objects with higher density and larger eccentricities of heliocentric orbits [103]. The above estimates are consistent with the fact that, in some cases, the clump that grew by accumulation of smaller objects could acquire the angular velocity necessary for the formation of a binary system.

Generally speaking, the clump angular momentum necessary for the formation of the Earth–Moon system could be gained by accumulating only small objects by a clump with the final mass $m_f > 0.15 m_E$. We believe, however, that the main contribution to the angular momentum of the parent clump was made by collisions of large clumps. Otherwise, Venus and Mars could have been formed together with large satellites (which is not the case) if their parent clusters had acquired sufficient angular momentum. Probably, unlike Earth, the clumps that formed the embryos of other terrestrial planets in the course of contraction did not collide with massive clumps at this stage. If this scenario is true, then the clump that gave rise to Mars's embryo did not have a large angular momentum, and only small satellites Phobos and Deimos could form during its contraction, although another mechanism for their appearance can also be envisaged. The angular momenta of the clumps that gave rise to the embryos of Mercury and Venus were insufficient, even for the formation of small satellites.

We emphasize that objects ejected from Earth's embryo in its collisions with other objects were more likely to become part of a large embryo of the Moon than to stick to similar smaller-mass objects. This contributed to the formation and growth of a larger Earth satellite than would have been if formed only from matter ejected from Earth. The presence of an Earth satellite formed during the contraction of the clump can explain the absence of Venus's satellites. Various planetesimals fell on Venus and on Earth with approximately the same distributions of masses and velocities. Under these collisions, matter was also ejected from the surface of Venus, but no satellite was formed from it.

This approach, which suggests that the initial embryos of the Moon and the Earth could form from a common parental clump, differs substantially from earlier studies [206–212] in which the Moon embryo was assumed to form and grow mainly due to Earth's crust material ejected from Earth's embryo during its numerous collisions with protoplanetary disk bodies. The main difference implemented in Ipatov's model [221] is that the initial embryo of the Moon was formed not from the substance ejected from Earth's embryo but from the same clump that Earth originated from, and the further growth of the Earth and Moon embryos formed under the contraction of the parent clump was similar to the multi-impact model. The matter that entered the Moon embryo could be ejected from Earth during numerous collisions of planetesimals and other smaller bodies with Earth, and not only during the $\sim 20$ major collisions, as was considered, for example, in [212].

A fundamentally important question regarding the origin of the Earth–Moon system is the cause of the differences in the iron composition of these bodies and the role of migration processes in that difference. Assuming that the fraction of iron in the initial embryo of the Moon and in planetesimals was 0.33, and the respective fraction of iron in Earth's crust and in the Moon, according to modern data, is 0.05 and 0.08, and using the relation $0.05 k_E + 0.33(1 - k_E) = 0.08$, we can estimate the fraction $k_E$ of Earth's crust substance in the composition of the Moon as $\sim 0.9$. Therefore, to explain such an iron content of the Moon, we should assume that the fraction of matter ejected from Earth's embryo and deposited on the Moon embryo was almost an order of magnitude greater than the total mass of planetesimals that fell directly on the Moon embryo and the initial mass of the Moon embryo formed from the parental clump, assuming that that embryo contained the same fraction of iron as the planetesimals [221, 226]. As we can see, the smaller the estimate of the fraction of Earth's crust matter in the Moon becomes, the more the Moon embryo, formed during the contraction of the clump, was depleted in iron and the greater its mass. As already noted, most matter that entered the Moon embryo could have been ejected from Earth during its numerous collisions with planetesimals and other bodies. We note that the ratio of the probability of a planetesimal collision with Earth to the probability of its collision with the Moon was less than the ratio of the Earth and Moon masses [146, 186]. Therefore, if all collisions of planetesimals with the Earth and the Moon had ended in mergers, the Moon growth rate would have been higher than that of the Earth. For a more accurate comparison of the growth rates for the Earth and Moon embryos, one should use the results of modeling planetesimal collisions with these embryos. We note that, due to the lower mass (and hence weaker gravitational field) of the Moon embryo compared with Earth's embryo, some high-speed collisions of planetesimals with the Moon could lead not to mergers but, conversely, to the ejection of matter from the Moon embryo surface and even to a decrease in its mass.

The models of formation of satellite systems that we propose impose some restrictions on the times of existence of rarefied clumps. For circular heliocentric orbits whose difference in the semi-major axes $a$ is equal to the Hill radius $r_{Ho}$, the ratio of the heliocentric periods of two clumps is about $1 + 1.5 r_{Ha}$, where $r_{Ha} = r_{Ho}/a$. In this case, the angle, with the apex placed at the Sun, between directions to the two clumps changes by $2\pi 1.5 r_{Ha} n_r$ rad in $n_r$ revolutions of the clumps around the Sun. We assume that the collision of clumps occurs when the semi-major axes of their orbits differ by $r_{Ho}$, and the initial angle, with the apex at the Sun, between the directions to the clumps is equal to $\pi$ rad. Then, the collision occurs after about $(3 r_{Ha})^{-1}$ revolutions. With a mass of $0.01 m_E$, the corresponding time between collisions is approximately $10^{8/3}/3 \approx 155$ revolutions. In other words, the collision of clumps that gave rise to the parental clump of the Earth–Moon system embryos could occur within about 100 years after their formation.

In [224], the number of collisions was studied for clumps with masses $m_0 = 10^{-7} m_E$ (e.g., solid bodies with the diameter $d_s = 100$ km and density $\rho \approx 1.15$ g cm$^{-3}$) moving in the same plane in the disk, with the ratio of the distances from the disk edges to the Sun $a_{rat} = 1.67$ (for example, the disk between 30 and 50 AU from the Sun) and the minimum distance $a_{min}$ from the Sun. If the surface density is the same over the entire disk, then the number of planetesimals at a distance from the Sun in the range from $a - 2r_H$ to $a + 2r_H$ is $N_{mH} = 8 N r_{Ha}(a/a_{min})^2/(a_{rat}^2 - 1)$, where $N$ is the number of clumps in the disk and $r_{Ha}$ and $a_{rat}$ are dimensionless quantities. The number was $N = 10^7$ for $m_0 = 10^{-7} m_E$ and the total disk mass $M_\Sigma$ equal to $m_E$ (the mass of the Earth). In this case, $N_{mH}$ ranges from $2.5 \times 10^3$ to $6.9 \times 10^3$ for the near and far edges of the disk. The average number $N_c$ of collisions of the Hill spheres of the considered clumps with other clumps in $n_r$ revolutions around the Sun can be estimated as $1.5 r_{Ha} n_r N_{mH}$, if we assume that collisions can occur when the semi-major axes of the orbits of converging clusters differ by no more than $2 r_{Ha}$ and the initial angle $\Delta\varphi_0$ between the cluster directions with the apex at the Sun is equal to $\pi$ radians. For $r_{Ha} \approx 4.6 \times 10^{-5}$ and the possibility of collisions of clumps with semi-major axes differing by $2 r_{Ho}$, the average number $N_{cl} = N_c/n_r$ of clump collisions per revolution around the Sun is 0.2 for $N_{mH} \approx 3 \times 10^3$ and 0.4 for $N_{mH} \approx 6 \times 10^3$. That is, in $3 \times 10^3$ revolutions, the fraction of collisions is approximately 20% of the number of initial clumps. The fraction of binary systems in the population of minor planets is assumed to be 0.3 for cold classical TNOs and 0.1 for all other TNOs. In the above model of the formation of satellite systems of TNOs during collisions of clumps, clumps should shrink several-fold over a time of about $3 \times 10^3$ revolutions around the Sun (760 thousand years at 40 AU). Estimates for this model are given for $\Delta\varphi_0 = \pi$. Earlier collisions occurred at smaller values of $\Delta\varphi_0$. Therefore, such a model gives an estimate of the clump contraction times from above. It was shown in [225] that the satellite system formation model based on the collision of two clumps is consistent with observations according to which about 40% of binary objects discovered in the trans-Neptunian belt have a negative angular momentum relative to their centers of mass.

## 4. Migration of planetesimals and planetary embryos in the feeding zone of the giant planets

The first calculations of the evolution of disks composed of hundreds of gravitating solid bodies that merge during collisions, for the final stages of accretion in the zone of the giant planets, date back to the 1980s [227–237]. Initial disks were assumed to include planetesimals and giant planet embryos whose initial orbits were close to the orbits of modern planets. Ipatov also modeled disks without planetary embryos and showed that the total mass of planetesimals ejected into hyperbolic orbits is an order of magnitude greater than the total mass of planetesimals that merged with planets. In the models in [234–237], the merger of bodies occurred when they approached distances of 4 to 8 radii of the bodies; when taking the gravitational influence of planets into account, spheres smaller than their spheres of influence were used, and the mutual gravitational influence of planetesimals was not taken into account. Therefore, the ejection of bodies into hyperbolic orbits turned out to be significantly suppressed compared with that obtained by Ipatov [228–230]. In addition, it was found in the models referred to above that, while the semi-major axis of Jupiter's orbit decreased, the semi-major axes of the orbits of the other giant planets mainly increased. The results of numerical simulations also suggest that, in the process of accretion of the giant planets, more ice and rocky material could enter Jupiter's core and shell than occurs with the other planets.

The idea of the formation of Uranus's and Neptune's embryos near the orbit of Saturn was first proposed in [175, 238]. Studying the composition of Uranus and Neptune, the authors of these papers concluded that the embryos of these planets acquired hydrogen shells with a mass of about $(1–1.5) m_E$ in Jupiter's and Saturn's growth zones even before the dissipation of gas from the protoplanetary disk occurred. Calculations in [145, 176, 239] showed that, under the influence of planetesimals that migrated from Uranus's and Neptune's feeding zones to Jupiter, the nearly formed Uranus and Neptune could migrate from Saturn's orbit to their modern orbits in 10 Myr, constantly moving in weakly eccentric orbits. Gravitational interactions were taken into account by the method of spheres. The eccentricities of Uranus's and Neptune's orbits remained small all this time. Later, similar calculations using the symplectic integration method were carried out in [240].

A model of the evolution of a disk that initially consisted of the terrestrial planets, Jupiter, Saturn, 750 identical bodies at a distance $R$ from 8 to 32 AU from the Sun with the total mass equal to $150 m_E$, and 150 smaller bodies at a distance $2 < R < 4$ AU was studied by Ipatov [103, 145]. In the course of evolution, the smaller bodies were swept out of the asteroid belt, and individual massive bodies acquired semi-major axes of highly eccentric orbits at $R < 2$ AU. Such bodies completely penetrated the feeding zone of the terrestrial planets. Similar results were obtained in modeling the evolution of disks with the same initial bodies but with the present-day Jupiter and Saturn masses and with the Uranus and Neptune embryo masses equal to $10 m_E$ moving in almost circular initial orbits (right-hand plots in Fig. 1). The initial values of the semi-major axes of the orbits of these planets were, respectively, 5.5, 6.5, 8, and 10 AU. In the course of evolution, Uranus and Neptune acquired orbits close to the present-day ones. We note that the results of calculations of such a migration [145, 176, 239] were published long before the first studies based on the above-mentioned Nice model appeared [241–243], and, unlike in the Nice model, the migration of Uranus's and Neptune's embryos occurred without resonances with the giant planets. Also, the total mass of planetesimals in Uranus's and Neptune's feeding zones was greater; in the variants considered, it ranged from $135 m_E$ to $180 m_E$, and more than 80% of planetesimals were ejected into hyperbolic orbits. It was concluded in [103] that a disk of planetesimals with the total mass equal to $100 m_E$ suffices for the migration of Uranus's and Neptune's embryos into present-day orbits. This mass decreases if the semi-major axes of the initial orbits of Uranus's and Neptune's embryos are taken larger than in the calculations (where they were 8 and 10 AU).

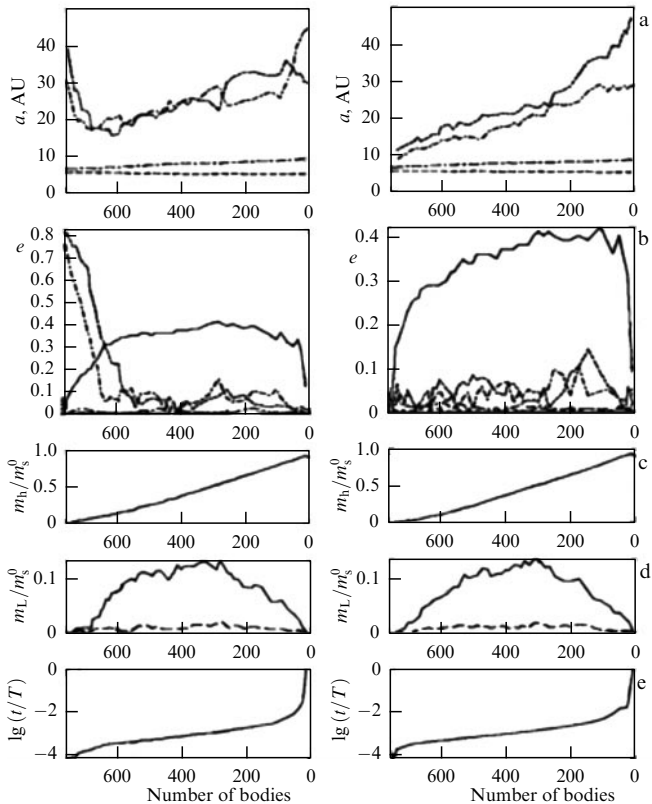

**Figure 1.** Dependences on the number of bodies in the disk for semi-major axes (a) and eccentricities of the orbits of giant planet embryos (dashed lines in Fig. b) and the mean eccentricity of the orbits of bodies (solid line in Fig. b), ratios of the total mass of bodies ejected into hyperbolic orbits to their initial total mass (c), ratios of the total mass of bodies with orbital semi-major axes greater than 49.5 AU to their initial total mass (d), and ratios of the current time to the total disk evolution time (e). Cases of large, 0.75–0.82 (left), and small, 0.02 (right), initial eccentricities of Uranus's and Neptune's embryo orbits are shown. (According to [103, 145].)

The main changes in the orbit parameters of the giant planet embryos occurred over a period no longer than 10 Myr, although individual bodies could fall on these embryos billions of years later. If most of the disk mass was due to small bodies, then the migration times of the planet embryos could be longer than with the initial values of the body masses chosen as $0.2 m_E$ in the calculations. In addition to calculations of the migration of the giant planets that were initially in circular or almost circular orbits, the case of large (0.75–0.82) initial eccentricities of the orbits of the massive ($10 m_E$) embryos of Uranus and Neptune was also considered (left-hand plots in Fig. 1) [103, 145]. In this model, the eccentri-cities of the embryo orbits decreased during the interactions of the embryos with planetesimals, and the orbits could transform into the modern orbits of Uranus and Neptune if the initial perihelia of their orbits lay beyond Saturn's orbit; at smaller perihelion distances, on the other hand, these embryos were in most cases ejected into hyperbolic orbits. But the acquisition by Uranus's and Neptune's embryos of such eccentric orbits with perihelia lying outside the orbit of Saturn is unlikely.

The total mass of bodies going beyond the orbit of Neptune could reach several ten Earth masses. The fact that Neptune has the smallest eccentricity among the giant planets may be due to the masses of the largest planetesimals in its feeding zone being less than in the zones of other giant planets. The dynamical lifetimes of most planetesimals inside the orbit of Neptune are less than 100 Myr. Therefore, the end of intense bombardment of Earth, which took place 4.5–3.8 billion years ago, could correspond in the most part to objects that came from eccentric orbits, mainly residing outside Neptune's orbit, or from the zone of the outer asteroid belt.

We briefly discuss the issue of the growth rate of the giant planets. In studying the migration of planetesimals initially located at different distances from the Sun, estimates were also obtained for the fraction of planetesimals that collided with the giant planets. The evolution of planetesimal orbits under the influence of planets was modeled by numerical integration of the equations of motion. The probability of a collision with Uranus or Neptune for a planetesimal initially located outside Jupiter's orbit did not exceed 0.015 and was no more than a few thousandths in most variants of calculations. Therefore, with the total mass of planetesimals beyond Saturn's orbit being less than $200 m_E$, the massive embryos of Uranus and Neptune that migrated from the zone in the vicinity of Saturn's orbit to the modern orbits increased their masses by no more than $2 m_E$. The probability of a planetesimal collision with Jupiter in most variants of calculations did not exceed 0.05 and was several times less for Saturn. According to the estimates in [244–247], the mass of the silicate component is $(15–20) m_E$ for Jupiter and is greater for Saturn. Along with silicates, planetesimals in Uranus's and Neptune's feeding zones also contained ice. Therefore, with the total mass of planetesimals beyond the orbit of Saturn being less than $200 m_E$, the increase in the mass of Jupiter's silicate component due to such planetesimals probably did not exceed several Earth masses. Hence, we can conclude that the total mass of planetesimals beyond the orbit of Saturn being not greater than $200 m_E$ is consistent with the composition of the giant planets.

With the ratio of the mass of dust made of rocky matter and ice to the mass of gas equal to 0.015 [248–250], the $200 m_E$ mass of planetesimals corresponds to a disk mass equal to $0.04 M_S$. Approximately the same values of the protoplanetary disk mass in the range of $(0.04–0.10) M_S$ were obtained in other studies [4, 34, 115, 250]. In a number of papers [167–170], the Grand Tack model was studied, where Jupiter moves in the asteroid belt zone while being formed, as we discussed in Section 3.2.

Of great interest is the question of how Jupiter was formed. A detailed review of relevant papers is given in [251]. The simulation results show that, with the formation of planetesimals at a distance of 5.2 AU, Jupiter's embryo could reach the mass of $3 m_E$ in 0.1 Myr, but the planetesimals in the vicinity of its orbit were then practically exhausted.

With the growth of Jupiter's embryo via the formation of much smaller ($\sim$ 1 cm–1 m) solids in the gas, its mass could increase to $10 m_E$ in just several ten thousand years. When Jupiter's embryo and the gas in its immediate vicinity reached some critical values of mass, the stage of rapid gas accretion began, during which Jupiter's embryo mass increased several-fold in a time of $\sim 0.1$ Myr. In some models, the time of Jupiter's mass growth from zero to approximately the present value was about 2 Myr [251]. However, already upon reaching a mass of about $10 m_E$ in a time of $\sim 0.1$ Myr, Jupiter's embryo was able to increase the orbital eccentricities of bodies from its feeding zone such that they could reach the feeding zone of the terrestrial planets at perihelia. As the masses of planetesimals increased and the gas density in Jupiter's feeding zone decreased, the mutual gravitational influence of the planetesimals increased the possibility for some planetesimals to start crossing Jupiter's orbit and then migrate to orbits with small perihelion distances.

A review of studies of secular perturbations of the planetary orbit elements was already given in a classic book by Subbotin [252] (Chapter XVIII, Section 7). In [253, 254], the evolution of the orbits of the four giant planets and of Pluto was studied in the respective intervals of $\pm 100$ and $\pm 50$ Myr. In [253], the bounds for the semi-major axes $a$, eccentricities $e$, and inclinations $i$ of the orbits of these planets were calculated and changes in Pluto's orbit elements were plotted. In [255], the equations of motion were integrated for the 8 planets on a 30 Myr interval and plots of the secular changes in $e$ and $i$ for the terrestrial planets were presented. Ipatov [256] obtained plots of changes in orbital elements of the planets due to their mutual gravitational influence. The integration was carried out over an interval of 20 Myr into the past. In particular, the respective orbital eccentricities were found to have changed from the present-day values to 0.07, 0.06, 0.127, 0.09, 0.076, and 0.02 for Venus, Earth, Mars, Saturn, Uranus, and Neptune. This means that they could differ significantly from the present-day values. At the same time, for Uranus's, Neptune's, and Pluto's orbits, the respective range of changes in the semi-major axes was 0.23, 0.41, and 1 AU.

## 5. Formation of the asteroid and trans-Neptunian belts

A number of authors [120, 129, 257, 258] developed a model according to which the rearrangement of resonances associated with a change in the semi-major axis of Jupiter's orbit, along with the influence of planetesimals from the feeding zones of the giant planets, could be one of the reasons for the sweeping of bodies from the main asteroid belt. Studies of the evolution of disks that initially consisted of the terrestrial planets, Jupiter, Saturn, 250 planetesimals with a total mass $m_{js}^0 \sim 10 m_E$ and semi-major axes of their initial orbits from 5 to 10 AU, and 250 'asteroids' with semi-major axes of initial orbits from 2 to 4 AU [103, 145] showed that, in this scenario, the semi-major axes of the orbits of Jupiter, $a_j$, and Saturn, $a_s$, respectively decreased by $0.005 m_{js}^0/m_E$ AU and $0.01 m_{js}^0/m_E$ AU. In calculations with the eight planets and initial bodies in Uranus's and Neptune's feeding zones, $a_j$ decreased by $0.005 m_{un}^0/m_E$ AU, and $a_s$ increased by $(0.01$–$0.03) m_{un}^0/m_E$ AU, where $m_{un}^0$ is the total initial mass of bodies in Uranus's and Neptune's feeding zones. For $m_{un}^0/m_E \geqslant 100$, the shifting resonances overlapped with a significant part of the asteroid belt. The dependences of the change in $a_j$ on $m_{un}^0$ and $m_{js}^0$ were approximately the same, which means that the changes in $a_j$ were mainly dependent on the total mass of planetesimals in the feeding zone of the giant planets rather than on the mass distribution over distances in this zone. In the course of several million years, $a_j$ first decreased by $0.005 m_{js}^0/m_E$ AU due to the ejection of bodies by Jupiter from Jupiter's and Saturn's feeding zones, and then more slowly decreased by $0.005 m_{un}^0/m_E$ AU due to the ejection of bodies initially located beyond the orbit of Saturn. The positions of the resonances changed, and some bodies penetrated the asteroid belt zone and the feeding zone of the terrestrial planets. At some stages of the evolution, the orbits of about 1% of the bodies initially located in Uranus's and Neptune's feeding zones crossed Earth's orbit. Values of $m_{js}^0$ much smaller than the actual total mass of planetesimals in Jupiter's and Saturn's zones were assumed in the calculations in order to see the effect exerted by lower-mass planetesimals on 'asteroids.' An increase in the average eccentricities of the 'asteroid' orbits to values not less than in the present-day asteroid belt was obtained, with most of the 'asteroids' ejected into hyperbolic orbits; 5% of the 'asteroids' fell on Venus and 2.5% on Earth. In this model, Mercury and Mars even left the Solar System, which was apparently because the masses used in the calculations (equal to $0.04 m_E$) were much greater than the average masses of real planetesimals in Jupiter's and Saturn's feeding zones.

The evolution of similar disks consisting of asteroids and massive bodies in Jupiter's and Saturn's zones was also considered in [259]. In [151], the influence of the residual gas in the disk was studied and it was noted that, with the current eccentricity of Jupiter's orbit, most of the asteroid belt bodies were swept out due to the motion of secular resonance. In [260], it was assumed that the initial asteroid belt could have even been empty.

The formation of TNOs with diameters of 100–1000 km from planetesimals with diameters of 1–10 km was studied in [261–269]. In these models, the TNO formation process took place at small eccentricities (usually $e \sim 0.001$) and for a massive belt (tens of $m_E$). A runaway growth of objects in 100 Myr for $e = 0.001$ and in 700–1000 Myr for $e = 0.01$ was obtained in [266]. These times are longer than the formation time of massive Jupiter, which, as we have seen, does not exceed tens of millions of years. In other calculations that take the gravitational influence of the giant planets into account [103], the maximum TNO eccentricities always exceeded 0.05 in a period of 20 Myr. Obviously, deceleration in gas could reduce the planetesimal orbit eccentricities, and the gravitational influence of the forming giants could be less than that of modern planets. It is unlikely, however, that, in the presence of the gravitational influence of the forming giant planets and the mutual gravitational influence of planetesimals, small eccentricities could persist for the time necessary for the formation of TNOs more than 100 km in diameter from planetesimals 1 to 10 km in diameter. It is more probable [270] that TNOs with diameter $d \geqslant 100$ km formed in the $a > 30$ AU zone by contraction of large rarefied clumps rather than accumulating small solids. According to [271], planetesimals with a diameter of several hundred kilometers in the zone of giant planets and large asteroids could form similarly. Some smaller objects could be fragments of these large objects, while other small objects could form directly by contraction of clumps. Even if the masses of the initial clumps into which the dust disk fragmented were approximately the same at some distance from the Sun, the processes of their merger and contraction gave rise to a rather arbitrary distribution of the resulting solids over mass [2, 24]. At a certain stage, oligarchic growth (run-away accretion) begins such that large planetesimals grow much faster than others due to the absorption of smaller ones. A similar effect can be observed when rarefied clumps merge.

The total mass of planetesimals entering the trans-Neptunian belt from the feeding zones of the giant planets could reach several ten $m_E$. These planetesimals increased the orbital eccentricities of local TNOs, whose total initial mass could exceed $10 m_E$, and swept most of these bodies out of this zone. A small fraction of these planetesimals may have remained beyond Neptune's orbit in highly eccentric orbits. Such a mechanism of TNO formation in highly eccentric orbits and the formation of 'local' TNOs in weakly eccentric orbits from matter located outside Neptune's orbit [230], as well as first estimates of the gravitational interaction of TNOs, had been discussed even before the 1992 discovery of the first TNO beyond the orbit of Neptune [272, 273]. It was shown that, over the past 4 Gyr, several percent of TNOs could change their semi-major axes $a$ by more than 1 AU due to gravitational interactions with other TNOs. We note that even small changes in the TNO orbit elements, occurring as a result of mutual gravitational influence and collisions, can lead to significant changes in the TNO orbit elements under the gravitational influence of the planets [274]. The swinging of TNO orbits by three very large ($1.5 m_E$) 'planetesimals' was modeled in [275]. Currently, TNOs moving in highly eccentric orbits with perihelion distances $q$ exceeding 40 AU are known (extended scattered disk objects), while typical scattered disk objects have perihelion distances $q \sim 35$–38 AU. In our opinion, many such TNOs with large $q$ could be former planetesimals from the feeding zone of giant planets.

It can be assumed that the population of trans-Neptunian bodies with diameter $d > 100$ km has not changed significantly in the time equal to the age of the Solar System because, at relatively low collision velocities in this belt, even collisions of large bodies with $d \sim 100 - 150$ km are unlikely and, according to estimates, about 5% of such bodies participated in such collisions [268]. A body with a mass of about $\sim 100$ times less than a 100-km object is not capable of destroying the latter, but can change its orbital velocity by several m s$^{-1}$ and its semi-major axis by $\sim 0.2\%$ ($\sim 0.1$ AU). Such events can generally be frequent enough to provide some flow of bodies the size of Chiron. The semi-major axes of the fragments formed in the collision then differ by 0.1 to 1 AU from the semi-major axes of the parent bodies.

## 6. Volumes of water and volatiles delivered to the terrestrial planets

The migration of small Solar System bodies, reflecting its dynamical properties, is related to a number of planet formation processes. According to current ideas shared by most researchers in the field of planetary cosmogony, the delivery of water and volatiles from the outer to the inner regions of the Solar System had a decisive effect on the evolution of the terrestrial planets, primarily on the Earth, making it suitable for life. This question is relevant, because the Earth and the terrestrial planets were formed in the high-temperature ($\sim 1000$ K) zone of the protoplanetary disk, where water and volatiles were not retained but accumulated beyond the ice line at a distance $R > 3$ AU. Studying the role of migration processes is therefore important for understanding the key problem of the origin of life, which is basic for astrobiology [1, 2, 276]. One way or another, despite a number of constraints, explaining the presence of water and volatiles on the Earth, which are mainly concentrated in the oceans and the atmosphere, by the migration of bodies from the outer regions of the Solar System allows circumventing the complications associated with the formation of the terrestrial planets in the high-temperature zone of the protoplanetary disk.

Endogenous and exogenous sources of water are considered the main potential mechanisms for the formation of Earth's oceans and presumed ancient oceans on Venus and Mars. Both mechanisms have certain limitations and can contribute jointly to solving this problem. Endogenous mechanisms are investigated by geochemical studies of the intrusive substance of magmatic melts of Earth's lithosphere, while studies of exogenous mechanisms are based on computer simulations. Both approaches allow reconstructing the geological history of Earth, accounting for a range of dynamical processes that have occurred throughout the history of the Solar System.

Endogenous sources of water could include direct adsorption of hydrogen from nebular gas into magma melts, followed by the reaction of $H_2$ with FeO, which could increase the D/H ratio in Earth's oceans 2 to 9 times [277], and the accumulation of water by the protoplanetary disk particles before the start of gas dissipation in the inner part of the young Solar System [278, 279]. The idea of a high water content in the mantle is supported by a number of studies, including laboratory analyses of olivine in Archean komatiite-basaltic associations (ultramafic lavas in Earth's green belts) formed during melting under extreme conditions at the boundary of Earth's upper mantle [280]. These results indicate the melting of the mantle at a temperature of 1630 K and a partial water content $\sim 0.5\%$, extrapolating which to the entire volume of the mantle corresponds to several Earth oceans. In [278], the volume of water in the minerals of silicate Earth is estimated at 5 to 6 (up to 50) volumes of Earth oceans. It was noted in [281] that deep mantle water could have been acquired as a result of water adsorption on fractal particles during Earth's accretion period and has a low D/H ratio.

Exogenous sources of water and other volatiles are the migration processes from the outer to the inner regions of the Solar System. They could include the migration of bodies from the outer part of the main asteroid belt [150, 169, 282–286] and the migration of planetesimals from beyond Jupiter's orbit [282, 287–294]. The migration of planetesimals from the 6–9.5-AU zone was considered within the Grand Tack model [170]. In these scenarios, the probability of collision of bodies with the Earth and other terrestrial planets and the mass of delivered water and other volatiles were estimated. According to [278], the fraction of bodies from beyond Jupiter's orbit did not exceed $\sim 50\%$ of the water delivered to the Earth.

A number of authors advocated the hypothesis that most of the water came to the Earth from the outer asteroid belt zone. For example, according to [283], several embryos that came from this zone at the final stage of Earth's formation could deliver an amount of water to Earth that was an order of magnitude greater than the current value. Such embryos could be the size of Mars [169]. But this hypothesis has not been corroborated. The key argument against such an ample asteroid source of water on the Earth was data on the isotope composition of osmium (Os) in the primary upper mantle of Earth, which turns out to be closer to ordinary anhydrous chondrites than to hydrous carbonaceous C-chondrites [278]. It is unlikely that the main source of water could be the outer asteroid belt rather than the zone of the giant planets and endogenous sources, because, if C-asteroids had come from the feeding zones of the giant planets [260], then the D/H ratio in Earth's oceans would have been similar to that in bodies directly arriving on the Earth from these zones.

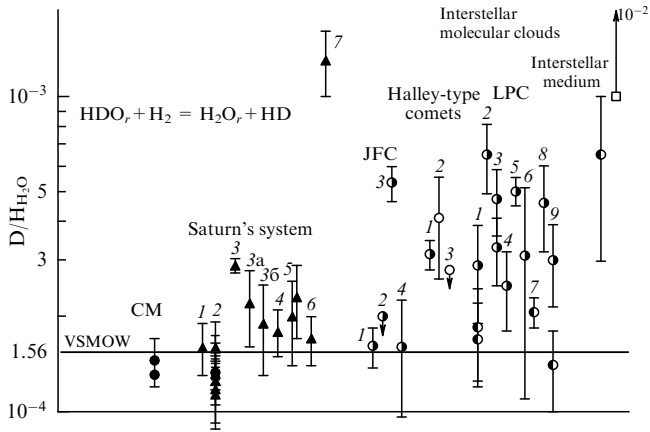

**Figure 2.** Hydrogen isotope ratios (D/H$_{H_2O}$) in water molecules measured in objects of the outer Solar System in comparison with data for carbonaceous CM chondrites and Earth's Vienna Standard Mean Ocean Water (VSMOW). Data are presented for the Saturn system, short-period Jupiter family comets (JFCs), Halley-type comets, and long-period comets (LPCs). (According to [295].)

A more probable source of water on the Earth could be planetesimals, akin to comets, in the feeding zones of the forming giant planets. The vast majority of these bodies, according to current ideas, were ejected into hyperbolic orbits and the periphery of the Solar System in the course of evolution and formed the Oort cloud, but many entered the zone of the terrestrial planets. At present, their main pools in the planetary system are the Jupiter- and Neptune-family comets. These bodies, whose icy matrix is abundant in water and volatiles, could become a source of heterogeneous accretion at the final stage of the formation of the Earth and other terrestrial planets. A well-known limitation of the model is the difference between the D/H ratio [295] in comets (except for a few comets, in particular, 103P/Hartley2) and the standard $D/H = 1.5576 \times 10^{-4}$ value in Earth's oceans (standard mean ocean water, SMOW), which lies in the range of $\sim (2–4) \times 10^{-4}$ (Fig. 2). This limitation is removed, however, if we assume that there were several sources of exogenous water on the Earth, along with comets and CI and CM chondrites.

A number of researchers believe that the difference between D/H values was determined at the earliest stages of the planetary system formation. According to [296], most of the water in the oceans was delivered by bodies formed in Jupiter's zone, where vapor from the inner Solar System condensed onto icy interstellar particles before they were accreted onto large bodies. It is believed in [278] that the measured D/H and Ar/O ratios in the coma and tails of comets do not faithfully represent the composition of comet nucleus material. It was also shown in [297] that the D/H water ratio is different for bodies formed at different distances from the Sun: low for the hot inner disk, increasing with distance from the Sun, and then decreasing again. According to [260], C-type asteroids were formed at a distance of 5 to 20 AU from the Sun and acquired their present-day orbits when gas was still present in this zone, on a time scale of 3–5 Myr [298]. In [299], the cause of the low D/H ratio for water in Earth's oceans is associated with the deuterium-to-tritium ratio affected by cosmic radiation, which acted by implanting a neutron in deuterium on dust particles in the disk.

As we can see, due to a number of factors, the D/H ratio could be different in many planetesimals from the feeding zones of the giant planets, in asteroids from the inner and outer belts, and in comets. In addition, as already noted, some contribution from an endogenous source can be assumed, and the study of lavas [281] showed that deep mantle water has a rather low D/H ratio. In other words, as noted in [186], the water in Earth's oceans (and hence the modern SMOW D/H ratio) could be the result of mixing water from several sources, with large and small D/H ratios.

In a number of studies [186, 288–294] based on numerical simulations, quantitative estimates were obtained for the delivery of water and volatiles to the Earth and the terrestrial planets. The authors studied the migration of tens of thousands of small bodies (Jupiter-family comets or planetesimals) and dust particles that originated from such bodies. The gravitational influence of seven planets (from Venus through Neptune) was taken into account. A symplectic integrator was used to integrate the equations of motion [300]. In particular, the orbit evolution was studied for > 30,000 bodies with initial orbits close to those of Jupiter-family comets, Halley-type comets, long-period comets, and asteroids in the 3/1 and 5/2 resonances with Jupiter, together with > 20,000 dust particles [289–294]. Integration continued until all bodies or particles reached the distance 2000 AU from the Sun or collided with the Sun. Based on the orbital elements of migrating bodies or particles obtained with a certain time step in the course of the dynamical lifetime, the probabilities of their collisions with planets were studied. The mean value of the probability $p_E$ of colliding with the Earth exceeded $4 \times 10^{-6}$ for a Jupiter-family comet and was $p_E = 2 \times 10^{-6}$ for a planetesimal from Jupiter's and Saturn's zones [186]. Half of that value is obtained if the gravitational influence of the planets is taken into account in calculations by the method of spheres [301].

Figures 3 and 4 show the values of $p_E \times 10^6$ for various distances from the Sun. Each value is derived from the evolution of the orbits of several hundred (up to 2000) planetesimals under the gravitational influence of the planets. In each run of the calculations, the initial values of the semi-major axes of the orbits of 250 planetesimals ranged from $a_{min}$ to $a_{min} + d_a$ [AU], the initial eccentricities were equal to $e_0$, and the initial inclinations were $0.5e_0$ rad. For each pair of $a_{min}$ and $e_0$ values, several (up to 8) calculation runs were launched. For $a_{min} \geqslant 3.6$ AU (except at $a_{min} = 4.2$ AU), the considered time interval was such that no more than a few percent of the initial bodies remained in elliptical orbits at the end of evolution. For $a_{min} \leqslant 3.6$ AU, the time interval reached the lifetime of the Solar System. Figure 3 shows the $p_E \times 10^6$ values for $d_a = 2.5$ AU and $a_{min}$ from 2.5 to 4.0 AU, and Fig. 4, for $d_a = 0.1$ AU and $a_{min}$ from 3.0 to 4.9 AU. For one among several thousand planetesimals, the probability of a collision with the Earth could be greater than the total probability for thousands of other planetesimals. If Fig. 3 were drawn with the data for such planetesimals excluded, then, instead of two $p_E$ peaks at 7.5 and 10 AU, there would be values close to the $p_E$ values for the neighboring $a_{min}$. Nevertheless, we see from Figs 3 and 4 that the $p_E$ values tend to decrease as $a_{min}$ increases; in Fig. 4, for $3.2 \leqslant a_{min} \leqslant 4.1$ AU, the values of $p_E$ are on average substantially greater than those for $a_{min} \geqslant 4.2$ AU.

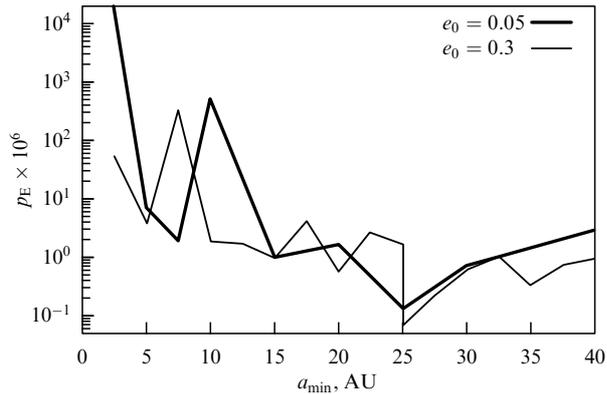

**Figure 3.** Probability $p_E$ of a planetesimal colliding with the Earth as a function of $a_{min}$. In each run of calculations, initial values of semi-major axes of planetesimal orbits changed from $a_{min}$ to $a_{min} + 2.5$ AU. Different lines correspond to initial orbit eccentricities equal to 0.05 or 0.3.

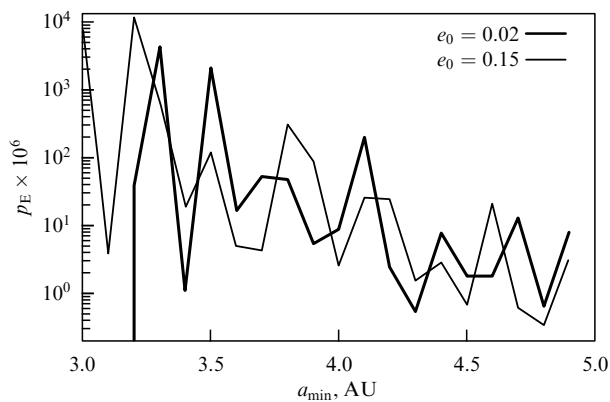

**Figure 4.** Probability $p_E$ of a planetesimal colliding with the Earth as a function of $a_{min}$. In each run of calculations, initial values of semi-major axes of planetesimal orbits changed from $a_{min}$ to $a_{min} + 0.1$ AU. Different lines correspond to initial orbital eccentricities equal to 0.02 or 0.15.

Using the value $p_E = 2 \times 10^{-6}$ and the estimate of the total mass of $\sim 100 m_E$ of planetesimals from Jupiter's and Saturn's feeding zones given in Section 4, we obtain the total mass of bodies that fell on the Earth at $2 \times 10^{-4} m_E$. Approximately the same mass of bodies could be acquired by the Earth from beyond the orbit of Saturn and the outer asteroid belt. Assuming that the ice of water and other volatiles accounted for about half this mass, we obtain the result that the total mass of water delivered to the Earth from beyond the ice line was $\sim 2 \times 10^{-4} m_E$, which corresponds to the mass of Earth's oceans ($2 \times 10^{24}$ g) or is slightly less if we assume that the fraction of ice in planetesimals was $\sim 1/3$ [186].

A significant fraction of water could have been delivered to Earth's embryo when its mass was much less than the modern Earth mass. The results of numerical simulation show that, when the embryo grew to half the modern mass of Earth, it received $\sim 30\%$ of the total volume of water delivered to the Earth from Jupiter's and Saturn's feeding zones [186]. The volume of water delivered to Venus, per unit mass of the planet, turned out to be approximately the same as for Earth, and that to Mars, approximately two to three times greater. These estimates show that the terrestrial planets could receive an amount of water and volatiles comparable to or even greater than that received by the Earth due to planetesimals from the feeding zone of the giant planets, which testifies in favor of the hypothesis positing ancient oceans on Mars and Venus [302]. The mass of water in bodies that came from beyond Jupiter's orbit and collided with the Moon could be less than that for bodies that collided with the Earth, but no more than by a factor of 20 [186]. However, the fraction of water evaporated during collisions of bodies with the Moon was greater than during collisions with the Earth.

Naturally, these values are estimates. This applies primarily to calculations of the probability of collisions between planets and bodies from beyond Jupiter's orbit. In particular, the results of simulations with 250 bodies showed that the $p_E$ values calculated in different variants with the same initial values of the semi-major axes and eccentricities could differ by a factor of 100 or more. Thus, in calculations of the migration of planetesimals with semi-major axes of the initial orbits from 3 to 5 AU, the $p_E$ values beyond Jupiter's orbit varied from less than $10^{-6}$ to about $10^{-3}$, although they were typically confined to between $10^{-6}$ and $10^{-5}$. Therefore, for better statistics and improved accuracy of simulation results, as many initial bodies as possible must be used and, accordingly, large computing power is needed. However, this remark does not affect the main conclusion about the important role of the exogenous source of water and volatiles in the evolution of the Earth and can only somewhat change the above quantitative estimates.

## 7. Migration of bodies from the asteroid and trans-Neptunian belts to the Earth's orbit and the problem of the asteroid–comet hazard

The main asteroid belt, the trans-Neptunian belt (Kuiper belt), and the Oort cloud are considered to be the main sources of near-Earth objects (NEOs) with a perihelion distance $q < 1.33$ AU. The fraction of close encounters of active comets with the Earth among encounters of all bodies is about 0.01, but 'extinct' comets, whose nuclei, like asteroids, show no activity, can be much more numerous than active comets.

The problem of the asteroid-comet hazard (ACH) is connected with the problem of asteroids, comets and meteoroids approaching the Earth. It attracts increasing attention as a potential source of threats to Earth's civilization. Among the NEOs, the greatest danger is posed by the three main groups of near-Earth asteroids of the Amor, Apollo, and Aten groups. Bodies from the Amor group come close to the Earth's orbit and those from the Apollo and Aten groups cross it, their semi-major axes being respectively greater or less than 1 AU. There is also a small group of asteroids of the Atira group whose orbits lie inside the Earth's orbit (Fig. 5). Some NEOs reach kilometer sizes, and their collision with the Earth could cause a global catastrophe, as has happened more than once in the geological history of our planet (see, e.g., [1]). ACH questions are discussed in many papers (see, e.g., [271, 288, 303–309]). The degree of hazard depending on the body size and the expected rate of events is determined according to the so-called Torino scale (Fig. 6), which, however, does not have an official international status. The fall of the Chelyabinsk meteorite in 2013 [307, 310] testifies to the serious consequences of the fall of even a relatively small ($\sim 20$ m) body in a densely populated area. Figure 6 allows estimating for bodies of various sizes the probability of collisions with the Earth and the kinetic energy released in this case on a scale from 1 (minimal damage) to 10 (global catastrophe). Obviously, the larger the size of the bodies that pose a real threat in the ACH context, the lower the probability.

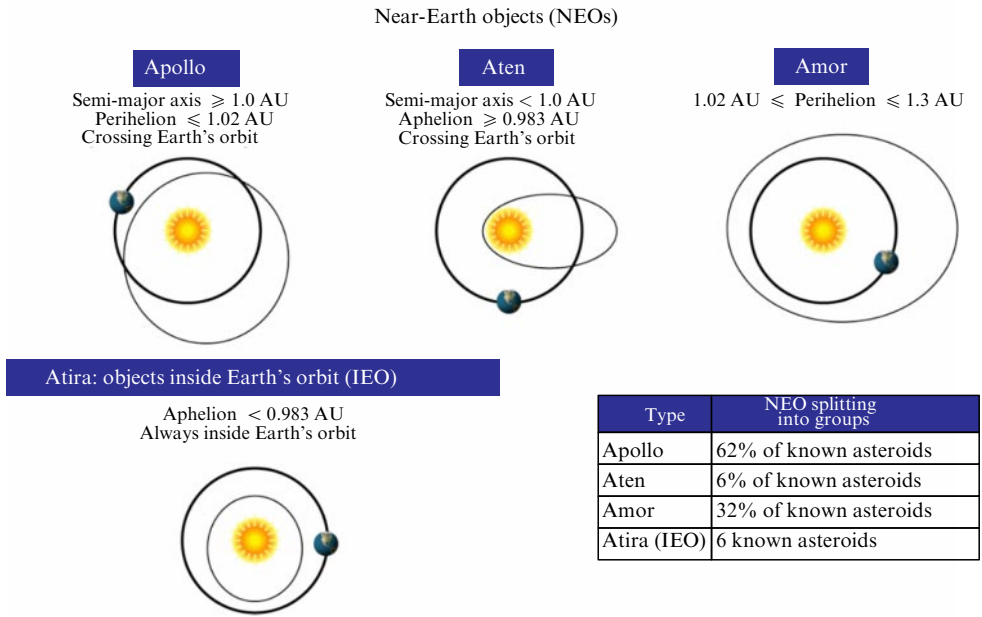

**Figure 5.** Near-Earth objects. (Source: James Green, NASA.)

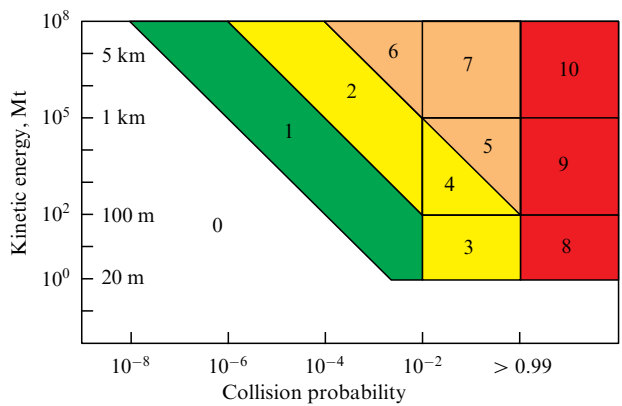

**Figure 6.** Torino scale used to determine the probability (expectation value) of a collision with Earth for a cosmic body (asteroid or comet) depending on the kinetic energy of the collision, expressed in TNT megatonnes, and the diameter of the body at a typical collision speed. The hazard score ranges from 1 (minimum) to 10 (maximum, threatening a global catastrophe). (Source: Wikipedia.)

Many researchers (see, e.g., [308, 311]) believe that asteroids are the main source of NEOs, while comets are more dangerous due to their sudden appearance. During the evolution of the Solar System, the migration of small bodies from the Kuiper belt led to the replenishment of NEOs. According to [270, 312], almost all NEOs came from the trans-Neptunian belt, while in [313] at least half the NEOs were assumed to be former short-period comets. The cometary nature of NEOs was also advocated in [314]. The cometary origin of some asteroids is indicated by the spectral characteristics of meteorites, which differ from those of asteroids [315]. In [316], about 40 active and 800 extinct comets of the Jupiter family with a diameter of more than 1 km and period $P < 20$ years, crossing Earth's orbit, and about 140 to 270 active Neptune-family Halley-type comets ($20 < P < 200$ years) were assumed to exist. According to the estimates in that paper, active and extinct comets are responsible for about 20% of craters greater than 20 km in diameter on the Earth.

The most effective NEO sources are resonances in the main asteroid belt. NEOs can be replenished from the inner part of the asteroid belt (in particular, due to the $v_5$, $v_6$, and $v_{16}$ secular resonances) [317–322] and can also come from Kirkwood gaps [313, 323–326]. Small asteroids can enter into resonances due to the Yarkovsky effect and as a result of mutual collisions. When bodies establish a resonance, they can greatly increase the eccentricities of their orbits, reaching the orbits of Mars and Earth, and also leave the resonance zones due to close encounters with these planets. In all likelihood, Kirkwood gaps in the main asteroid belt — regions corresponding to minima in the distribution of orbit semi-major axes $a$ — were formed in that way. This hypothesis about the origin of the gaps was put forward based on the ratios of the heliocentric orbital periods of Jupiter and an asteroid found to be 3:1 [327, 328], 5:2 [324, 329, 330], and also 2:1 or 7:3. For a number of hypothetical asteroids in the 5:2, 3:1, and 2:1 ratios, with quasiperiodic changes in eccentricities $e$, an increase in $e$ was obtained from 0.15 to, respectively, 0.75, 0.45, and 0.35 [331]. Studies (including by numerical integration of the equations of motion) of the evolution of the orbits of hypothetical asteroids in the vicinity of the 5:2 resonance with Jupiter were carried out in [331–336]. In [337, 338] for the 3:1 resonance and in [324–326, 334] for the 5:2 resonance, it was found that the boundaries of the regions of the initial $a$ and $e$ values that allow reaching the orbit of Mars are close to the boundaries of the corresponding Kirkwood gaps.

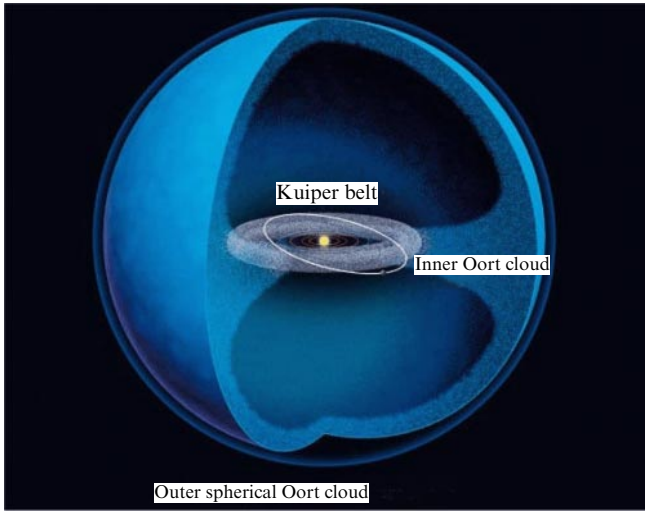

**Figure 7.** Location of the Oort Cloud and the Kuiper Belt. (Source: http://galspace.spb.ru/index376.html.)

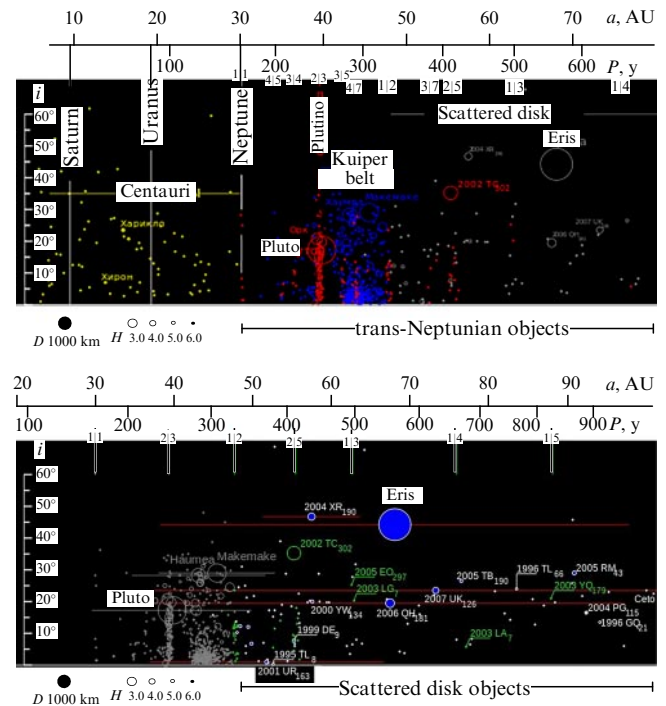

**Figure 8.** Comparative sizes and inclinations of some TNO orbits. (Source: https://ru.wikipedia.org/wiki/.)

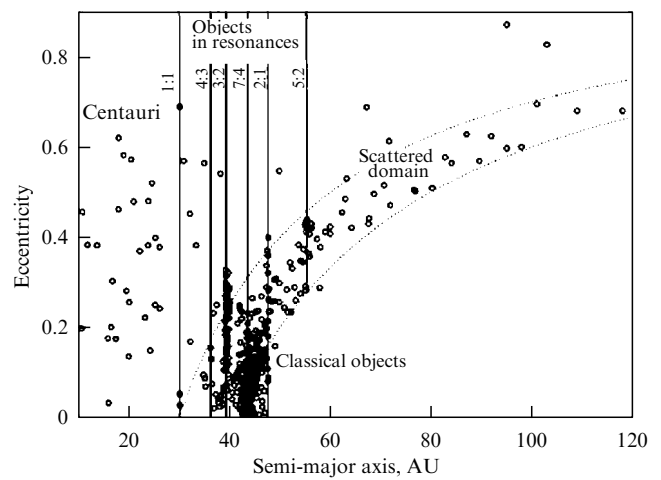

**Figure 9.** Semi-major axes and eccentricities of TNO orbits. (Source: Scott Sheppard, Carnegie Institution for Science.)

It was shown in [339, 340] that, in the presence of mean motion resonances with Mars (3:5, 7:12, 4:7, 5:9, 7:13, and 1:2) and with Jupiter (7:2 and 10:3), and a three-body resonance among Jupiter, Saturn, and an asteroid, or among Mars, Jupiter, and an asteroid, matter can be delivered to Mars and Earth. Because even small changes in the semi-major axis of an asteroid can bring it into the resonance region, the number of bodies that delivered matter to Earth was significant, and collisions with them pose a threat in the modern era.

Based on model calculations [308], it is believed that most of the NEOs came from the inner part of the main asteroid 3:1 resonance region with Jupiter, 40% from the $\nu_6$ secular resonance, and 13% from an intermediate source of orbits crossing the orbit of Mars, while 16% are former Jupiter-family comets. It also follows from the model that asteroids from the outer part of the belt ($a > 2.8$ AU) passing to NEO orbits on average spend only 140 thousand years there, which is 16 times less than in the case of the 3:1 resonance and much less than for other zones within the main asteroid belt. To estimate the fraction of each of the four sources mentioned above, the evolution of the orbits of several real asteroids and test Jupiter-family comets was modeled, and the characteristic elements of the NEO orbits obtained for different ratios were compared with the actual distribution. It was shown that there may be a large number of unobserved extinct comets whose aphelia lie inside Jupiter's orbit and which also serve as a source of NEOs. The median lifetime of NEOs is 10 Myr [308, 341], with more than half of them falling on the Sun, 10–20% falling on planets (mainly on Venus and Earth), and 15% being ejected from the Solar System. Among the bodies that came from the zones of the giant planets, the fraction of those that cross Earth's orbit is an order of magnitude greater than the fraction of bodies residing in orbits that cross only Mars's orbit, and they usually have $e > 0.6$. These results indicate that most asteroids of the Amor group did not come from the zones of the giant planets but from the asteroid belt.

NEO sources, along with the main asteroid belt, also include TNOs (scattered disk and Kuiper belt objects) and comets from the Oort cloud (Figs 7–9). They are closely related to each other genetically and evolutionarily. The connection of short-period comets with the trans-Neptunian belt was established in [342], the migration of TNOs to Neptune's orbit [343–345] and the migration of bodies from the orbit of Neptune to Jupiter's orbit [345, 346] were studied, and the probability of the transition of objects crossing Jupiter's orbit into Encke-type orbits was found to be less than 0.0023 [347]. The cosmic ages of meteorites, which establish the time interval from the moment of their breakaway from the parent body or the complete breakup of the latter, are spread over a wide range of $10^6$–$10^9$ years [348].

Asteroid–comet bodies in the trans-Neptunian region, including the main family of dwarf planets typified by Pluto, are concentrated in the Kuiper belt ($30 \leqslant a \leqslant 50$ AU) and in the scattered disk ($a > 50$ AU). The total mass of bodies in the Kuiper belt was estimated as $0.06 m_E$–$0.25 m_E$ [349], with a later estimate of $0.02 m_E$ [350]. For belt objects with $a \leqslant 50$ AU, the mean eccentricity and mean inclination are $e_{av} \approx 0.1$ and $i_{av} \approx 8°$. The number of such objects with diameter $d \geqslant 1$ km is estimated as $5 \times 10^9$ [351] to $10^{10}$–$10^{11}$ [349, 352], and with diameter $d \geqslant 100$ km, of the order of 70,000 [346, 349]. Scattered disk objects have $e_{av} \approx 0.5$ and $i_{av} \approx 16°$, and the total mass of bodies moving in highly eccentric orbits between 40 and 200 AU is estimated as $0.5 m_E$ in [353] and $0.05 m_E$ in [354]. The number of scattered disk objects with $d \geqslant 100$ km is about 30,000 [354]. The probability that a trans-Neptunian body with $a < 50$ AU leaves the belt in a year is $(3-5) \times 10^{-11}$ [343], although some objects could be more likely to reach the orbit of Neptune due to mutual gravitational influence [273].

The evolution of orbits of small bodies under the gravitational influence of planets was studied in detail by numerical integration of the equations of motion in [103, 271, 274, 355, 356]. In modeling the evolution of the orbits of about a hundred Kuiper belt objects, it was found that the perihelia of the orbits of two such objects decreased in the course of evolution to 1 AU in 25 and 64 Myr. The mean time of the body motion in the orbit of a Jupiter-crossing object (JCO) is 200 thousand years, the fraction of JCOs that reach Earth's orbit during their lifetime is 0.2, and the mean time during which JCOs cross Earth's orbit is about 5000 years. Based on these results, it was concluded that, with $10^{10}$ Kuiper belt objects 1 km in size and 750 objects 1 km in size that cross Earth's orbit (ECOs), the number of present-day JCOs coming from the trans-Neptunian belt is $N_J = 30,000$, and about 170 former TNOs (about 20% of the ECOs) cross the Earth's orbit. Objects that cross the Earth's orbit and have lost cometary activity stay relatively far from the Earth most of the time and are more difficult to observe than typical ECOs.

The number of NEOs and their orbital parameters (for example, mean inclinations, which are greater than those for objects in the main asteroid belt) are difficult to explain if only asteroid sources are considered, as was noted in [158]. In our opinion, the mean orbital inclination of the NEOs that arrived from the main asteroid belt may be the same as for all NEOs ($\sim 15°$) and may be greater than the mean orbital inclination in this belt ($\sim 10°$), because these objects could increase their orbital inclinations when they were in resonance.

The migration of comets from Neptune's orbit to the interior of the Solar System was first studied by Kazimirchak-Polonskaya [345]. The evolution of the orbits of real short-period comets under the influence of planets over an interval of 10 Myr was simulated numerically by Levison and Duncan [346]. Calculations showed that 91% of comets were thrown into hyperbolic orbits, 1% and 0.1%, respectively, collided with Jupiter and Saturn, and 6% (including Encke's Comet) evaporated during close encounters with the Sun (Sungrazers). The median lifetime of comets turned out to be 0.45 Myr. In the model in [346], the evolution of the orbits of objects leaving the trans-Neptunian belt was studied. For such objects, the median lifetime was 45 Myr, during which 30% became visible comets (with $q < 2.5$ AU), 99.7% of them belonging to the Jupiter family. Comparing the observed distribution of the orbits of Jupiter-family comets with the distribution obtained in the calculations, the authors concluded that the physical lifetime of Jupiter-family comets is 12,000 years. According to their estimates, a Jupiter-family comet falls on Jupiter about once every 400 years, and on Earth once every 13 Myr. Twenty-five % of the objects considered were ejected into hyperbolic orbits, 68% acquired $a > 1000$ AU, and 1.5% collided with planets, including 30% with Neptune, 33% with Uranus, 13% with Saturn, and 23% with Jupiter. The median lifetime of Halley-type comets (with a period $20 < P < 200$ years), prior to their ejection into hyperbolic orbits, was estimated to be 1 Myr [357]. It can be concluded that TNOs are an excellent source of Jupiter-family comets and that they produce almost no Halley-type comets, which came mainly from the Oort cloud. According to [358], Jupiter-family comets are in resonance with Jupiter for a part of their lifetime. The velocities of collisions with the Moon for bodies that came from Jupiter's and Saturn's feeding zones were mainly in the range from 20 to 23 km s$^{-1}$, and with Earth, 3 km s$^{-1}$ higher [359].

The time of cometary activity has been estimated by many scientists; on average, it is about $10^3$–$10^4$ years (according to [360], from $3 \times 10^3$ to $3 \times 10^4$ years). Some comets that have lost their activity can continue moving for tens and even hundreds of millions of years in orbits that cross Earth's orbit. Therefore, the number of extinct comets can exceed the number of active comets by several orders of magnitude. During close flybys of planets, comets can break apart into several parts, like Comet Shoemaker–Levy 9 during its close approach to Jupiter in July 1992, two years before its collision with Jupiter [361, 362].

Asteroids and comet nuclei are responsible for the formation of craters on the surfaces of planets, their satellites, and small bodies themselves. It is believed that a relatively uniform size distribution of interplanetary impactors of mixed origin was established as early as 4 Gyr ago [363] and that the flow of crater-forming bodies has been approximately constant over the past 3 Gyr; about 4 Gyr ago, during the late heavy bombardment, it was 100–500 times more intense [309, 364]. The rate of NEO collisions with Earth, according to estimates in [365], is 2, 14, 24, and 30 times higher compared to Venus, Mars, the Moon, and Mercury, respectively. It was shown in [366] that collisions with asteroids probably dominated the formation of terrestrial craters with a diameter $D < 30$ km, while comet impacts were responsible for the formation of craters with diameter $D > 50$ km. This feature can be explained by the fact that TNOs can leave the trans-Neptunian belt almost without collisions, in contrast to the main asteroid belt bodies, which have experienced multiple collisions.

Data on lunar craters less than 100 m in diameter suggest that the flow of crater-forming bodies has been approximately constant over the last 100 Myr [309]. It was assumed in [367] that fragments of an asteroid destroyed about 160 Myr ago in the asteroid belt could have formed the Baptistina family, which then caused an increase in the flow of bombarding bodies. An analysis of the ages of Copernican period craters (less than 1.1 Gyr) led Mazrouei et al. [368] to the conclusion that the number of NEO collisions with the Moon per unit time increased 290 Myr ago by about a factor of 2.6. Estimates of the age of the craters were based on an analysis of the thermophysical properties of the material ejected during impacts, according to the Diviner radiometer on the LRO lunar satellite (USA). At the same time, a shortage of terrestrial craters aged between 300 and 650 Myr and their almost complete absence for a later age were noted. We remark that the assumption of a twofold increase in the number of craters over the past 300 Myr was made earlier in [369] based on the study of bright beams of ejecta from craters. Estimates for the age of ray craters on the far side of the Moon are less than 1 billion years.

The number of lunar craters $> 15$ km in diameter with an age $< 1.1$ Gyr was compared with estimates of the number of craters that could have been formed in 1.1 Gyr, assuming that the number of NEOs > 1 km in diameter and their orbital elements remained close to their modern values during that period [370]. The comparison was made for craters on the entire surface of the Moon with data for a region of the Ocean of Storms and maria on the visible side of the Moon. The probabilities of NEO collisions with the Moon and the dependences of crater diameters on the impactor diameters were used. The number of known Copernican craters with diameter $D \geqslant 15$ km per unit area of maria was noted to be at least twice that number for the rest of the lunar surface. The results are also consistent with the idea of an increase in the number of NEOs after a possible catastrophic destruction of large main belt asteroids that could have occurred during the past 300 Myr, but they do not prove that such an increase actually took place. In particular, they are consistent with the conclusion in [368] that the number of collisions of near-Earth asteroids with the Moon per unit time increased 290 Myr ago by a factor of 2.6. If the probability of a collision with Earth per year for an object crossing Earth's orbit is $10^{-8}$ (over long time intervals), estimates of the number of craters presented in [370] correspond to a model in which the number of 15-km Copernican craters per unit area for the entire surface of the Moon would be the same as for the mare region if the data [371] for $D < 30$ km were as complete as for $D > 30$ km. In that case, the rate of crater formation could be approximately constant over the last 1.1 Gyr. The dependences of the depths of lunar craters on the diameter were considered in [372]. It was noted that craters on lunar maria are deeper than on the continents, the crater diameters being less than 30 to 40 km. Having larger diameters, the continental craters are deeper.

## 8. Migration of dust in the Solar System and the formation of the zodiacal cloud

In addition to comets and asteroids, numerous meteoroids fall on the terrestrial planets. They are bodies intermediate in size between interplanetary dust and asteroids (less than a few ten meters in diameter). They, in particular, include the parent body of the Chelyabinsk meteorite mentioned above. The number of meteoroids grows exponentially as their size and mass decrease. It is estimated that about 98 to 99% of such bodies weighing less than 100 g in the vicinity of Earth are mainly of cometary origin.

Particles formed during asteroid collisions and ejected by comets during the sublimation of the icy core matrix are the main source of interplanetary dust. Their size ranges from nanometers to millimeters, and the lower threshold is at the upper bound of the size of molecular clusters. The millimeter–centimeter borderline is used as a convention to distinguish between dust particles and meteoroids. The dust particles attending sublimation from a comet nucleus form dust tori surrounding their comet orbits, which are periodically crossed by the Earth and cause meteor showers. The known meteor showers are directly related to the radiants of parent comets projecting onto constellations after which these meteor showers are named. The orbit of the Geminid meteor shower practically coincides with the orbit of the asteroid 3200 Phaeton [373].

The amount of material contained in dust particles and small meteoroids and falling daily into Earth's atmosphere ranges from 30 to 180 t, of which 32 t is accounted for by bodies smaller than 0.5 m, according to [374]. At the early stage of planetesimal formation, the main role was played by dust particles of micrometer and millimeter sizes; along with planetesimals, particles of that size and larger probably made a significant contribution to the formation of planets.

The migration of dust particles has been considered by a number of authors (see, e.g., [289, 375–386]). In our numerical models [289, 292, 294, 384, 385], in addition to the gravitational influence of the planets, other factors (radiation pressure, the Poynting–Robertson effect, solar wind) were also taken into account. The relative error at the integration step in the Bulirsch–Stoer algorithm was less than $10^{-8}$. The initial particle orbits were assumed to be the same as for 500 real asteroids and for various comets and planetesimals from Jupiter's and Saturn's feeding zones. The ratio $\beta$ of the radiation pressure force to the gravitational force was in the range of 0.0004–0.4. For silicates, according to [378, 379], such values of $\beta$ correspond to particle diameters from 1000 to 1 μm. The planets were considered material points, but the elements of the orbits obtained with a 20-year step were used to calculate (similarly to calculations for small bodies) the average probabilities $p$ of particle collisions during their lifetime with the planets and the mean times during which the perihelia of the particle orbits were less than the semimajor axes of a planet's orbit.

Obviously, the smaller the particles (and, accordingly, the larger $\beta$), the lower the probability of their collision with the Sun, because more particles are carried away by the solar wind, which means a lower probability of particle collisions with the terrestrial planets. The probability $p_E$ of a dust particle colliding with the Earth turned out to be maximum ($p_E \sim 0.001$–0.02) at particle diameters of $\sim 100$ μm. Such values could be orders of magnitude higher than the probability of a planetesimal colliding with a planet for the same initial orbits of planetesimals and dust grains [292, 294], which is explained by smaller typical orbital eccentricities than planetesimals have and lower relative velocities of such particles when approaching the planet. A large amount of matter, including water and volatiles, could have been delivered by dust particles to the feeding zone of the terrestrial planets immediately after the zone near Jupiter's and Saturn's orbits was cleared of gas but retained a large amount of dust. Peaks in the distribution of the asteroid dust particles over the semimajor axes of their orbits, which correspond to $n:(n+1)$ resonances with Earth and Venus and gaps associated with 1:1 resonances with these planets, are more pronounced for larger particles. The probability of a collision of a trans-Neptunian particle with diameter $d < 10$ μm with the Earth is only several times less than for an asteroid particle of the same size. Dust particles are subject to much less heating at the altitudes of entry into the atmosphere due to the deceleration in it. Therefore, compared with large bodies, dust particles are considered a more likely means of interplanetary and interstellar transport of complex organic molecules, including biogenic elements and compounds that are part of microorganisms (organogenic substances according to V I Vernadsky [289]). These ideas are consistent with the panspermia hypothesis: dust particles could play an important role in the origin of life on Earth, because they experience much less heating when entering the atmosphere at low incidence angles.

The nature of zodiacal light is directly related to the migration of dust particles. Zodiacal light can be seen from the Earth as a bright diffuse glow, in the shape of a triangle, in the west after dusk and in the east before dawn, whose brightness decreases with increasing elongation. It is caused by an interplanetary dust cloud that lies in the ecliptic plane and reflects sunlight. The first scientific explanation of this phenomenon was given back in 1683 by Dominic Cassini.

Based on the calculated positions and velocities of migrating dust particles that started from various small bodies (asteroids, comets, TNOs), changes in the zodiacal light spectrum, which generally coincides with the solar Fraunhofer spectrum, were considered in the case of scattering by dust particles at various values of the angle, with the apex placed on the Earth, between the directions to the Sun and to the dust particle. The shift and width of the characteristic Mg I line were determined. The calculated data [386] were compared with data from WHAM (Wisconsin H-Alpha Mapper) observations of Doppler shifts and the relevant linewidth in zodiacal light [387]. It was shown that the contributions to the zodiacal light made by the cometary particles formed inside and outside Jupiter's orbit, counting the trans-Neptunian particles, are roughly the same, and the contribution of each of these two components to the zodiacal light is approximately 1/3, with a possible deviation from 0.3 to 0.1–0.2. The fraction of asteroid dust is estimated as $\sim 0.3-0.5$. The contribution of particles generated by long-period and Halley-type comets does not exceed 0.1–0.15. The same conclusion can be drawn for particles ejected by Encke-type comets (with $e \sim 0.8$–0.9). The average eccentricities of the orbits of zodiacal particles located at 1–2 AU from the Sun, which fit the WHAM observations better, have values in the range of 0.2–0.5, with a more probable value around 0.3.

Most recently, it was suggested, based on measurement data from the Juno spacecraft (USA) that one of the sources of replenishment of the zodiac cloud dust could be the dissipation of particles from the Martian atmosphere during planetary-scale dust storms [388]. The authors support this hypothesis with computer simulation results.

## 9. Migration of planetesimals in exoplanetary systems

More than 5000 exoplanets have been discovered so far, most of which belong to planetary systems of their parent stars. The current state and key problems of exoplanet research are discussed in monographs [15, 17] and Marov and Shevchenko's review [16]. The topical issues discussed there include the possible habitability of exoplanets, and in particular criteria for the formation of natural conditions suitable for the origin of life. Among these criteria, by analogy with the Solar System, the processes of planetesimal migration in exoplanetary systems can be important because of their role in ensuring the presence of water and volatiles in habitable-zone exoplanets.

Of paramount interest are exoplanetary systems located at the closest distances from the Earth. These, first of all, include the star Proxima Centauri, the third stellar companion (C) in the Alpha Centauri system. It is a red dwarf of spectral type M, located at a distance of 1.302 pc from the Sun. The star has a mass of 0.1, a radius of 0.15, and a visual luminosity of 0.00005 in terms of solar values. The effective temperature of its surface is $\sim 3000$ K, which is half the solar temperature. This star has a planet comparable to the Earth in mass, Proxima Centauri b, whose orbital radius is 0.05 AU (8 times less than the orbital radius of Mercury) and the orbital period is 11.2 days. However, with a much lower luminosity of the star, this planet is in the habitable zone, which extends in radius from $\sim 0.042$ to $\sim 0.082$ AU. The other planets, Proxima Centauri c and d, are outside this zone.

When modeling the migration of planetesimal exocomets in this stellar system, the semimajor axis $a_c$ of the exoplanet c orbit ranged from 0.06 to 0.3 AU (up to 0.7 AU in some test calculations) in [389], and was set as $a_c = 1.489$ AU in [390, 391], in agreement with [392, 393]. In [390, 391], we used a symplectic integrator [300] for integration. In [390], two series of calculations were carried out for the MP model. In this model, the planets are regarded as material points (points masses) when integrating, as we discuss in the next paragraph. In the first series of calculations, the initial values of the semi-major axes of the orbits and masses of two exoplanets were chosen as $a_b = 0.0485$ AU, $a_c = 1.489$ AU, $m_b = 1.27 m_E$, and $m_c = 12 m_E$. For the exoplanet b orbit, the initial eccentricity $e_b$ and inclination $i_b$ were set equal to zero, and for exoplanet c, $e_c = 0$ or 0.1 and $i_c = e_c/2$ rad. In the second run of calculations, to take later observational data into account, the values $a_b = 0.04857$ AU, $e_b = 0.11$, $m_b = 1.17 m_E$, $a_c = 1.489$ AU, $e_c = 0.04$, $m_c = 7 m_E$, and $i_b = i_c = 0$ were chosen. In both runs of calculations, the exoplanet b and c densities were considered equal to the respective densities of Earth and Uranus. In each version of the calculations, the initial values of the semi-major axes of the orbits of 250 planetesimals ranged from $a_{min}$ to $a_{min} + 0.1$ AU, where $a_{min}$ was varied from 1.2 to 1.7 AU with a step equal to 0.1 AU. The initial eccentricities $e_0$ of the planetesimal orbits were 0 or 0.15 for the first run and $e_0 = 0.02$ or $e_0 = 0.15$ for the second run, and the initial inclinations of their orbits were equal to $e_0/2$ rad. In the C model [391], the planetesimals that collided with planets were excluded from subsequent integration. In these calculations, $a_{min}$ ranged from 0.9 to 2.2 AU, and the rest of the initial data were the same as for the second run of the MP calculations. The considered time interval was usually equal to several hundred Myr for the C model and at least 50 Myr for the MP model.

In the MP [390] model, planetesimals and exoplanets were regarded as material points, and their collisions were not simulated. The resulting arrays of orbital elements of planetesimals with a step of 100 years were used to calculate the probabilities of their collisions with exoplanets. The probabilities were calculated using a technique similar to that used in [146, 290–292], but with the suitable exoplanet and star masses. In the case where the collision probability reached 1 (no such cases were observed in calculations for the Solar System), this planetesimal was no longer taken into account in calculating the total probability of collision with an exoplanet. We also calculated the probability $p_d$ for a planetesimal that migrated from the feeding zone of exoplanet c to collide with exoplanet d ($a_d = 0.02895$ AU, $m_d = 0.29 m_E$, and $e_d = i_d = 0$), although this exoplanet was not considered when integrating the equations of motion.

The orbit of only one of several hundred planetesimals crossed the orbit of exoplanet b, but such a planetesimal collided with this exoplanet quite often. When considering thousands of planetesimals, the value of $p_b$ turned out to be greater than the probability of a planetesimal from the feeding zone of the giant planets in the Solar System colliding with the Earth. In the second run of MP calculations (involving 250 initial planetesimals in each case) with $e_0 = 0.02$, the total number of planetesimals was 4500, and in only 5 out of 16 cases were the resultant probabilities of collisions with exoplanet b nonzero. In two variants, $p_b = 0.004$ was obtained, while the average value of $p_b$ for one of the 4500 planetesimals was $4.7 \times 10^{-4}$, but this included two planetesimals with $p_b = 1$. Of the 16 variants of the second run of MP calculations, four cases yielded a nonzero value of $p_d$. For the 4500 planetesimals, the average probability of collision with exoplanet d was $p_d = 2.7 \times 10^{-4}$, but this included one planetesimal with $p_d = 1$. For $e_0 = 0.15$, nonzero values of $p_b$ were obtained in only three out of six variants of the second run of MP calculations with 250 initial planetesimals (with 1500 planetesimals in total). The mean value for 1500 planetesimals turned out to be $p_b = 2.0 \times 10^{-3}$ and $p_b = 1$ for three of them, and the mean value was $p_d = 2.0 \times 10^{-3}$, with $p_d = 1$ for three planetesimals. In the first run of MP calculations with $i_c = e_c = 0$ and $e_0 = 0.15$, the probabilities $p_c$ of a collision of a planetesimal initially located in the vicinity of exoplanet c with that exoplanet were $p_c = 0.06$–0.1. For $i_c = e_c/2 = 0.05$ and $e_0 = 0.15$, this was $p_c = 0.02$–0.04. In the second series of MP calculations, the value of $p_c$ was mainly in the range of 0.1–0.3, except in cases with $a_{min} = 1.4$ AU and $e_0 = 0.02$, where $p_c = 0.4$–0.8. After 20 Myr, the increase in $p_c$ was usually small, because few planetesimals were left in elliptical orbits by that time.

Calculations for the C model, in which planetesimals colliding with a planet were excluded from subsequent integration, were done in [391] for initial data similar to those for the second run of MP calculations. In the C model, the probability $p_b$ of a planetesimal from the feeding zone of planet c colliding with planet b was about half as much; on the contrary, the values of $p_c$ were on average almost twice as large as in the MP model. The C-model probability $p_b$ was estimated to be $2.0 \times 10^{-4}$ for $e_0 = 0.02$ and $10^{-3}$ for $e_0 = 0.15$. The total mass of planetesimals delivered from the feeding zone of planet c to planet b was $m_{c-b} = p_b m_{fzc}$, and the mass of water in these planetesimals was $m_{ice} = p_b k_{ice} m_{fzc}$, where $m_{fzc}$ is the total mass of planetesimals beyond the ice line that fell into the feeding zone of Proxima Centauri c and $k_{ice}$ is the water content in planetesimals. The ratio $p_{cej} = p_c/p_{ej}$ of the probability of a planetesimal colliding with planet c to the probability $p_{ej}$ of the planetesimal ejection into a hyperbolic orbit at $e_0 = 0.02$ and $e_0 = 0.15$ was in the respective ranges of 0.8–1.3 and 0.4–0.6 when calculating with the current planet c mass. This ratio was in the ranges of 1.3–1.5 and 0.5–0.6 for the planet c mass equal to half its modern value. An estimate based on the values of $p_{ej}$ obtained from the energy conservation law showed that the semi-major axis of the planet c orbit could decrease during its formation by at least a factor of 1.5. The calculations showed that planetesimals could collide with planet b, even if the mass of the planet c embryo was one tenth that of the modern planet c mass. Therefore, the total mass of planetesimals ejected into hyperbolic orbits by planet c could be about $(3.5-7)m_E$. For the mass of planet c equal to $7m_E$, we obtain the mass $m_{fzc}$ of planetesimals in the feeding zone of planet c possibly being at least $10m_E$ and $15m_E$ for the respective values $e_0 = 0.02$ and $e_0 = 0.15$. For $m_{fzc} = 10m_E$ and $p_b = 2 \times 10^{-4}$, we obtain the minimum estimate $m_{c-b} = 2 \times 10^{-3} m_E$ (for $e_0 = 0.02$) of the total mass of planetesimals delivered to planet b from the feeding zone of planet c. For $e_0 = 0.15$, $m_{fzc} = 15m_E$, and $p_b = 10^{-3}$, this bound was $m_{c-b} = 1.5 \times 10^{-2} m_E$. Large values of $e_0$ correspond to an increase in the eccentricities of planetesimal orbits due to their mutual gravitational influence. With large masses of planetesimals, the increase in the eccentricities of the planetesimal orbits could be greater. The estimates of $m_{c-b}$ made above for the modern mass of planet b could be smaller if the planet b mass was less than its modern mass at the time of the considered bombardment. However, it can be assumed that planet b formed faster than planet c, because planet b is much closer to the star than planet c. The amount of matter delivered to planet d could be slightly less than to planet b. As we showed in [186, 394], the probability of collision with Earth for a planetesimal from the feeding zone of the giant planets is of the order of $10^{-6}$–$10^{-5}$, i.e., much less than the values of $p_b$ and $p_d$. As we can see, the inflow of icy planetesimals to Proxima Centauri exoplanets b and d could be greater than the similar inflow to the Earth.

Some of the material of a planetesimal that collides with an exoplanet is ejected from the exoplanet. It was found in [385] that more than 50% of the impactor's water is lost if the planetesimal collides with the Earth at a speed exceeding the parabolic velocity by more than 1.4 times and the collision angle is greater than 30°. It was assumed in [396] that solids beyond the ice line should be $\sim 50\%$ water by mass. According to [62], the fraction of ice in comet 67P is in the range of 14–33%. In [397], it was assumed that, although the volume fraction of water in comet 67P and TNOs is about 20%, the bodies formed near the ice line contained more water than the TNOs did. In studies of the maximum water content resulting from late accretion on TRAPPIST-1 planets [398], it was assumed that late impactors contained 10% water by mass. Generalizing the above data, we can assume that the fraction of water in planetesimals in the feeding zone of Proxima Centauri c could be 10 to 50%. The mass of water delivered to Proxima Centauri b could exceed the mass of Earth's oceans.

After the formation of Proxima Centauri c, some planetesimals could continue to move in stable elliptical orbits inside its feeding zone, mostly cleared of planetesimals, although hundreds of millions of years have passed since the beginning of calculations. Such planetesimals typically moved in some resonances with planet c, for example, 1:1 (like Jupiter's Trojans), 5:4, and 3:4, and had low eccentricities. Some planetesimals that moved for a long time (1 to 2 Myr) along chaotic orbits were captured in 5:2 and 3:10 resonances with Proxima Centauri c and remained there for at least tens of millions of years [399].

The mixing of planetesimals in the TRAPPIST-1 exoplanetary system was studied in [400] at the late gasless stage of the formation of nearly formed planets. The TRAPPIST-1 system consists of a star with a mass equal to 0.0898 solar masses and seven planets located relatively close to each other. The motion of planetesimals under the gravitational influence of a star and seven planets (from b to h in the order of their distance from the star) was studied similarly to the C model analysis for the Proxima Centauri system. In each of the calculation versions, the initial orbits of planetesimals were in the vicinity of the orbit of one of the planets (the host planet) and had the same eccentricities, equal to $e_0$. No more than 3.2% of planetesimals were ejected into hyperbolic orbits. Usually, there was no ejection of planetesimals for the b–d disks. More than half the planetesimals of the b–g disks collided with planets in less than 1000 years, and even in 250 years for the b–d disks. The time of evolution of the b–h disks varied from 11 ky to 63 Myr. The fraction of planetesimals that collided with the host planet was 0.36–0.8 for $e_0 = 0.02$ and 0.22–0.75 for $e_0 = 0.15$. The fraction of planetesimal collisions with the host planet was typically smaller for disks that were more distant from the star. In each version of the calculations, there was at least one planet for which the number of planetesimal collisions exceeded 25% of the number of planetesimal collisions with the host planet. Planetesimals could collide with all planets for the d–h disks, and at least with the b–e planets for b–c disks. Therefore, the outer layers of neighboring planets in the TRAPPIST-1 system can include similar material if there were many planetesimals near their orbits at the late stages of planetary formation. For comparison, we concluded in Section 3.2 that, due to the mixing of planetesimals, such a scenario could to a certain degree affect the formation of the terrestrial planets.

## 10. Conclusion

The problems of small body migration are among the most important dynamical properties of the Solar System. They have played a key role in its formation and evolution. We extensively studied these properties using the results of numerical simulations of migration processes. We have discussed the issues of the evolution of the protoplanetary disk, the formation of planets, the formation of the asteroid and trans-Neptunian belts, and the role of planetesimal migration processes in the formation and growth of planetary embryos in the emerging Solar System. Models of isolated formation of the terrestrial planets and their formation models, taking into account the influence of the giant planets, the issues of migration of planetesimals and embryos of planets in the feeding zone of the giant planets, and the time scales of the corresponding processes have been discussed. It is shown that Earth and Venus could acquire more than half of their mass in 5 million years, and their outer layers could accumulate the same material from different parts of the feeding zone of these planets. At the final stages of the formation of the terrestrial planets, planetesimals initially located at a radial distance of 1.1 to 2.0 AU could enter the composition of Earth and Mars in a ratio not much different from the mass ratio of these planets.

Models of asteroid and comet migration from the asteroid belt and zones beyond the orbits of Jupiter and Neptune to the Earth and the terrestrial planets have been discussed. Asteroids and comets, enriched in water and volatiles, could make a decisive contribution to the formation of the hydrosphere and atmosphere of Earth, which emerged in the high-temperature zone of the protoplanetary disk, where volatiles are not retained. At the same time, the considered migration models of bodies approaching or crossing Earth's orbit (the bodies genetically and evolutionarily associated with all zones of the Solar System) contribute to the urgent problem of ACH for our planet.

We critically reviewed modern models of the origin of the Moon, including the popular mega-impact model, the multi-impact model of planetesimal collisions with Earth's embryo, and the model of the formation of Earth's and Moon's embryos as a result of contraction of a rarefied dust cluster. The best substantiated model is the one according to which the angular momentum of the clump necessary for the formation of the Earth–Moon system embryos was acquired during the collision of two initial clumps, and most of the matter that entered the Moon's embryo could be ejected from Earth during numerous collisions of planetesimals with it.

We also analyzed existing models of the migration of the giant planets in the early Solar System. Based on data from the analysis of the composition of the giant planets, it was shown previously that large embryos of Uranus and Neptune formed near Saturn's orbit. The results of numerical calculations indicate that such embryos could migrate to their modern orbits under the influence of gravitational interactions with planetesimals. Most of the planetesimals were then ejected into hyperbolic orbits, and the semimajor axis of Jupiter's orbit decreased (and in the Grand Tack model, increased after some time).

The migration of small bodies is of key importance in studying the possibility of the formation of favorable natural conditions for the origin of life not only in the Solar System but also in exoplanetary systems. Exogenous sources of water on the Earth could include the migration of bodies from the outer part of the main asteroid belt and the migration of planetesimals from beyond the orbit of Jupiter, the feeding zones of the giant planets, and trans-Neptunian space, including the Kuiper belt. The mass of water delivered to Earth from these sources could, according to estimates, be comparable to the volume of Earth's oceans. Per unit mass of the planet, this mass was almost the same for Venus and for Earth, and about two to three times greater for Mars. These estimates support the hypothesis of the possible existence of ancient oceans on Mars and Venus.

We also discussed the migration of dust in the Solar System and sources of the zodiacal cloud. It is shown that the contribution to the zodiacal light made by cometary particles formed inside and outside Jupiter's orbit (counting the trans-Neptunian particles) is approximately the same, and the contribution of each of these two components to zodiacal light is $\sim 1/3$, with a possible deviation of 0.1–0.3. The fraction of asteroid dust is $\sim 0.3 - 0.5$. The contribution of particles from long-period and Halley-type comets does not exceed 0.10–0.15. A similar conclusion can be drawn for particles generated by Encke-type comets. The mean eccentricities of the orbits of zodiacal particles located at 1–2 AU from the Sun have values in the range from 0.2 to 0.5.

The likely processes of planetesimal migration in exoplanetary systems have been discussed. Based on numerical modeling under a number of initial assumptions, it was concluded that the inflow of icy planetesimals to inner exoplanets in the Proxima Centauri system could be greater than the similar inflow to Earth.

**Acknowledgments.** Sections 1–7 (except Section 3.4) are devoted to studies of the processes of migration and formation of planets and small bodies of the Solar System and were financially supported by the Russian Foundation for Basic Research in the framework of research project 20-12-50142. Studies of the formation of the Earth–Moon system (Section 3.4) were supported by the Russian Science Foundation project 21-17-00120, https://rscf.ru/project/21-17-00120/. Studies of dust migration (Section 8) were carried out in the framework of the state assignment 0137-2019-0004 at the Vernadsky Institute for Geochemistry and Analytical Chemistry, Russian Academy of Sciences. Research on the migration of planetesimals in exoplanetary systems (Section 9) was supported by the Ministry of Science and Higher Education of the Russian Federation grant 075-15-2020-780, "Theoretical and experimental studies of the formation and evolution of extrasolar planetary systems and characteristics of exoplanets." The authors are grateful for this support. We are grateful to the anonymous referee for the useful comments that contributed to the improvement of the content of this paper.

## References

1. Marov M Ya *Kosmos. Ot Solnechnoi Sistemy Vglub' Vselennoi* (Space. From the Solar System Deep into the Universe) 2nd ed. (Moscow: Fizmatlit, 2018) Ch. 2

2. Marov M, in *Oxford Research Encyclopedias: Planetary Science* (Ed.-in-Chief P Read) (Oxford: Oxford Univ. Press, 2018) https://doi.org/10.1093/acrefore/9780190647926.013.2
3. Marov M Ya *The Fundamentals of Modern Astrophysics: A Survey of the Cosmos from the Home Planet to Space Frontiers* (New York: Springer, 2015)
4. Safronov V S *Evolution of the Protoplanetary Cloud and Formation of the Earth and the Planets* (Jerusalem: Israel Program for Scientific Translations, 1972); Translated from Russian: *Evolyutsiya Doplanetnogo Oblaka i Obrazovanie Zemli i Planet* (Moscow: Nauka, 1969)
5. Contopoulos G *Order and Chaos in Dynamical Astronomy* (Berlin: Springer, 2002)
6. Marov M Ya *Phys. Usp.* **48** 638 (2005); *Usp. Fiz. Nauk* **175** 668 (2005)
7. Armitage P J *Astrophys. J.* **665** 1381 (2007)
8. Boss A P et al. *Astrophys. J.* **686** L119 (2008)
9. Boss A P et al. *Astrophys. J.* **708** 1268 (2010)
10. Wasserburg G J, in *Protostars and Planets II* (Eds D C Black, M S Matthews) (Tucson, AZ: Univ. of Arizona Press, 1985) p. 703
11. Raymond S N, Barnes R, Kaib N A *Astrophys. J.* **644** 1223 (2006)
12. Raymond S N, Quinn T, Lunine J I *Icarus* **183** 265 (2006)
13. Raymond S N, Scalo J, Meadows V S *Astrophys. J.* **669** 606 (2007)
14. Wada K et al. *Astrophys. J.* **702** 1490 (2009)
15. Marov M Ya, Shevchenko I I *Ekzoplanety. Ekzoplanetologiya* (Exoplanets. Exoplanetology) (Moscow–Izhevsk: Inst. Komp'ut. Issled., 2017)
16. Marov M Ya, Shevchenko I I *Phys. Usp.* **63** 837 (2020); *Usp. Fiz. Nauk* **190** 897 (2020)
17. Marov M Ya, Shevchenko I I *Ekzoplanety. Fizika, Dinamika, Kosmogoniya* (Exoplanets. Physics, Dynamics, Cosmogony) (Moscow: Fizmatlit, 2022)
18. ALMA Partnership, Brogan C L et al. *Astrophys. J. Lett.* **808** L3 (2015)
19. Bjerkeli P, in *XXXth General Assembly of the Intern. Astronomical Union, Vienna, Austria, August 20–31, 2018*
20. Bjerkeli P et al. *Nature* **540** 406 (2016)
21. Jorgensen J K, in *XXXth General Assembly of the Intern. Astronomical Union, Vienna, Austria, August 20–31, 2018*
22. Kolesnichenko A V, Marov M Ya *Turbulentnost' i Samoorganizatsiya. Problemy Modelirovaniya Kosmicheskikh i Prirodnykh Sred* (Moscow: Binom. Laboratoriya Znanii, 2009)
23. Marov M Ya, Kolesnichenko A V *Turbulence and Self-organization. Modeling Astrophysical Objects* (New York: Springer, 2013); Translated from Russian: *Turbulentnost' i Samoorganizatsiya. Problemy Modelirovaniya Kosmicheskikh i Prirodnykh Sred* (Moscow: Binom. Laboratoriya Znanii, 2014)
24. Marov M Ya, Rusol A V *Astron. Lett.* **44** 474 (2018); *Pis'ma Astron. Zh.* **44** 517 (2018)
25. Wiltzius P *Phys. Rev. Lett.* **58** 710 (1987)
26. Chen T, in *XXXth General Assembly of the Intern. Astronomical Union, Vienna, Austria, August 20–31, 2018*
27. Smirnov B M *Fizika Fraktal'nykh Klasterov* (Physics of Fractal Clusters) (Moscow: Nauka, 1991)
28. Smirnov B M *Phys. Usp.* **34** 711 (1991); *Usp. Fiz. Nauk* **161** (8) 141 (1991)
29. Eletskii A V, Smirnov B M *Phys. Usp.* **34** 616 (1991); *Usp. Fiz. Nauk* **161** (7) 173 (1991)
30. Smirnov B M *Phys. Usp.* **34** 526 (1991); *Usp. Fiz. Nauk* **161** (6) 171 (1991)
31. Kolesnichenko A V, Marov M Ya *Solar Syst. Res.* **48** 354 (2014); *Astron. Vestn.* **48** 383 (2014)
32. Goldreich P, Ward W R *Astrophys. J.* **183** 1051 (1973)
33. Greenberg R et al. *Icarus* **59** 87 (1984)
34. Safronov V S, Vityazev A V *Proiskhozhdenie Solnechnoi Sistemy* (Origin of the Solar System) (Itogi Nauki i Tekhniki. Ser. Astronomiya (Results of Science and Technology. Ser. Astronomy), Vol. 24) (Moscow: VINITI, 1983) p. 5
35. Eneev T M, Kozlov N N *Solar Syst. Res.* **15** 59 (1981); *Astron. Vestn.* **15** 80 (1981)
36. Eneev T M, Kozlov N N *Adv. Space Res.* **1** 201 (1981)
37. Weidenschilling S J *Icarus* **165** 438 (2003)
38. Makalkin A B, Ziglina I N *Solar Syst. Res.* **38** 288 (2004); *Astron. Vestn.* **38** 330 (2004)
39. Johansen A et al. *Nature* **448** 1022 (2007)
40. Johansen A, Youdin A, Klahr H *Astrophys. J.* **697** 1269 (2009)
41. Johansen A, Youdin A, Mac Low M-M *Astrophys. J.* **704** L75 (2009)
42. Johansen A, Klahr H, Henning T *Astron. Astrophys.* **529** A62 (2011)
43. Johansen A, Youdin A N, Lithwick Y *Astron. Astrophys.* **537** A125 (2012)
44. Johansen A et al., in *Asteroids IV* (Space Science Series, Eds P Michel, F E DeMeo, W F Bottke) (Tucson, AZ: The Univ. of Arizona Press, 2015) p. 471
45. Johansen A et al. *Sci. Adv.* **1** 1500109 (2015)
46. Cuzzi J N, Hogan R C, Shariff K *Astrophys. J.* **687** 1432 (2008)
47. Cuzzi J N, Hogan R C, Bottke W F *Icarus* **208** 518 (2010)
48. Lyra W et al. *Astron. Astrophys.* **491** L41 (2008)
49. Lyra W et al. *Astron. Astrophys.* **497** 869 (2009)
50. Morbidelli A et al. *Icarus* **204** 558 (2009)
51. Chambers J E *Icarus* **208** 505 (2010)
52. Chiang E, Youdin A N *Annu. Rev. Earth Planet. Sci.* **38** 493 (2010)
53. Rein H, Lesur G, Leinhardt Z M *Astron. Astrophys.* **511** A69 (2010)
54. Youdin A N *Astrophys. J.* **731** 99 (2011)
55. Youdin A N, Kenyon S J, in *Planets, Stars and Stellar Systems* (Eds T D Oswalt, L M French, P Kalas) (Dordrecht: Springer, 2013) p. 1
56. Marov M Ya et al., in *Problemy Zarozhdeniya i Evolyutsii Biosfery* (Problems of the Origin and Evolution of the Biosphere) (Ed. E M Galimov) (Moscow: Krasand, 2013) p. 13
57. Jansson K W, Johansen A *Astron. Astrophys.* **570** A47 (2014)
58. Carrera D, Johansen A, Davies M B *Astron. Astrophys.* **579** A43 (2015)
59. Kretke K A, Levison H F *Icarus* **262** 9 (2015)
60. Jansson K W et al. *Astrophys. J.* **835** 109 (2017)
61. Ziglina I N, Makalkin A B *Solar Syst. Res.* **50** 408 (2016); *Astron. Vestn.* **50** 431 (2017)
62. Davidsson B J R et al. *Astron. Astrophys.* **592** A63 (2016)
63. Gundlach B et al. *Astron. Astrophys.* **583** A12 (2015)
64. Fulle M et al. *Mon. Not. R. Astron. Soc.* **462** S132 (2016)
65. Poulet F et al. *Mon. Not. R. Astron. Soc.* **462** S23 (2016)
66. Kolesnichenko A V, Marov M Ya *Solar Syst. Res.* **52** 44 (2018); *Astron. Vestn.* **52** 51 (2018)
67. Youdin A N, Goodman J *Astrophys. J.* **620** 459 (2005)
68. Kolesnichenko A V, Marov M Ya *Solar Syst. Res.* **43** 410 (2009); *Astron. Vestn.* **43** 424 (2009)
69. Johansen A, Youdin A, Klahr H *Astrophys. J.* **697** 1269 (2009)
70. Pan L et al. *Astrophys. J.* **740** 6 (2011)
71. Youdin A N, Shu F H *Astrophys. J.* **580** 494 (2002)
72. Armitage P J, astro-ph/0701485
73. Bai X-N, Stone J M *Astrophys. J.* **722** 1437 (2010)
74. Drążkowska J, Dullemond C P *Astron. Astrophys.* **572** A78 (2014)
75. Jacquet E, Balbus S, Latter H *Mon. Not. R. Astron. Soc.* **415** 3591 (2011)
76. Jansson K W et al. *Astrophys. J.* **835** 109 (2017)
77. Garaud P, Lin D N C *Astrophys. J.* **608** 1050 (2004)
78. Kolesnichenko A V, Marov M Ya *Solar Syst. Res.* **47** 80 (2013); *Astron. Vestn.* **47** 92 (2013)
79. Vityazev A V, Pechernikova G V, Safronov V S *Planety Zemnoi Gruppy: Proiskhozhdenie i Rannyaya Evolyutsiya* (Terrestrial Planets: Origin and Early Evolution) (Moscow: Nauka, 1990)
80. Myasnikov V P, Titarenko V I *Solar Syst. Res.* **23** 7 (1989); *Astron. Vestn.* **23** 14 (1989)
81. Myasnikov V P, Titarenko V I *Solar Syst. Res.* **23** 126 (1990); *Astron. Vestn.* **23** 207 (1989)
82. Cuzzi J N, Hogan R C, in *43rd Lunar and Planetary Science Conf., March 19–23, 2012, Woodlands, Texas* (LPI Contribution, No. 1659) (Houston, TX: Lunar and Planetary Institute, 2012) id. 2536
83. Gibbons P G, Rice W K M, Mamatsashvili G R *Mon. Not. R. Astron. Soc.* **426** 1444 (2012)
84. Nesvorny D, Youdin A N, Richardson D C *Astron. J.* **140** 785 (2010)
85. Jansson K W, Johansen A *Astron. Astrophys.* **570** A47 (2014)
86. Lambrechts M, Johansen A *Astron. Astrophys.* **544** A32 (2012)
87. Lambrechts M, Johansen A *Astron. Astrophys.* **572** A107 (2014)
88. Morbidelli A et al. *Icarus* **258** 418 (2015)

89. Morbidelli A *Astron. Astrophys.* **638** A1 (2020)
90. Valletta C, Helled R *Astrophys. J.* **900** 133 (2020)
91. Tanaka H, Takeuchi T, Ward W R *Astrophys. J.* **565** 1257 (2002)
92. Benítez-Llambay P et al. *Nature* **520** 63 (2015)
93. Ipatov S I, Preprint No. 102 (Moscow: Institute of Applied Mathematics of the Academy of Sciences of the USSR, 1981)
94. Beletsky V V *Essays on the Motion of Celestial Bodies* (Basel: Birkhäuser Verlag, 2001); Translated from Russian: *Ocherki o Dvizhenii Kosmicheskikh Tel* (Moscow: Nauka, 1977)
95. Beletskii V V *Regulyarnye i Khaoticheskie Dvizheniya Tverdykh Tel* (Moscow: Inst. Komp'yut. Issled., RKhD, 2007)
96. Amelin Y et al., in *37th Annual Lunar and Planetary Science Conf., March 13–17, 2006, League City, Texas* (Houston, TX: Lunar and Planetary Institute, 2006) id. 1970
97. Bouvier A et al. *Geochim. Cosmochim. Acta* **71** 1583 (2007)
98. Bouvier A, Wadhwa M, in *40th Lunar and Planetary Science Conf., March 23–27, 2009, Woodlands, Texas* (Houston, TX: Lunar and Planetary Institute, 2009) ID 2184
99. Connelly J N et al. *Science* **338** 651 (2012)
100. MacPherson G J, Krot A N *Meteorit. Planet. Sci.* **49** 1250 (2014)
101. Connelly J N et al. *Science* **338** 651 (2012)
102. Wetherill G W *Annu. Rev. Astron. Astrophys.* **18** 77 (1980)
103. Ipatov S I *Migratsiya Nebesnykh Tel v Solnechnoi Sisteme* (Migration of Celestial Bodies in the Solar System) (Moscow: Editorial URSS, 2000) https://doi.org/10.17513/np.451
104. Vityazev A V *Izv. Akad. Nauk SSSR. Fiz. Zemli* (8) 52 (1991)
105. Vityazev A V, Pechernikova G V *Sov. Astron.* **25** 494 (1981); *Astron. Zh.* **58** 869 (1981)
106. Vityazev A V, Pechernikova G V, Safronov V S *Sov. Astron.* **22** 60 (1981); *Astron. Zh.* **55** 107 (1978)
107. Gurevich L E, Lebedinskii A I *Izv. Akad. Nauk SSSR. Ser. Fiz.* **11** 765 (1950)
108. Ziglina I N *Solar Syst. Res.* **25** 526 (1991); *Astron. Vestn.* **25** 703 (1991)
109. Ziglina I N *Solar Syst. Res.* **29** 1 (1995); *Astron. Vestn.* **29** 3 (1995)
110. Ziglina I N, Safronov V S *Sov. Astron.* **20** 244 (1976); *Astron Zh.* **53** 429 (1976)
111. Ipatov S I *Sov. Astron.* **32** 560 (1988); *Astron. Zh.* **65** 1075 (1988)
112. Levin B Yu *Sov. Astron. Lett.* **4** 54 (1978); *Pis'ma Astron. Zh.* **4** 102 (1978)
113. Pechernikova G V *Sov. Astron.* **18** 778 (1975); *Astron. Zh.* **51** 1305 (1974)
114. Pechernikova G V, Vityazev A V *Sov. Astron.* **24** 460 (1980); *Astron. Zh.* **57** 799 (1980)
115. Ruzmaikina T V, Maeva S V *Astron. Vestn.* **20** 212 (1986)
116. Safronov V S *Astron. Zh.* **31** 499 (1954)
117. Safronov V S, in *Voprosy Kosmogonii* (Problems of Cosmogony) (Exec. Ed. B V Kukarkin) (Moscow: Izd. Akad. Nauk SSSR, 1960) p. 121
118. Safronov V S *Proiskhozhdenie Zemli* (Origin of the Earth) (Moscow: Znanie, 1987)
119. Safronov V S *Astron. Vestn.* **28** (6) 3 (1994)
120. Safronov V S, Ziglina I N *Astron. Vestn.* **25** 190 (1991)
121. Shmidt O Yu *Dokl. Akad. Nauk SSSR* **46** 392 (1945)
122. Shmidt O Yu *A Theory of Earth's Origin*; *Four Lectures* (Moscow: Foreign Languages Publ. House, 1958); Translated from Russian: *Chetyre Lektsii o Teorii Proiskhozhdeniya Zemli* (Moscow: Izd. Akad. Nauk SSSR, 1957)
123. Greenberg R et al. *Icarus* **94** 98 (1991)
124. Levin B J *The Moon and the Planets* **19** 289 (1978)
125. Lissauer J J *Icarus* **69** 249 (1987)
126. Lissauer J J *Annu. Rev. Astron. Astrophys.* **31** 129 (1993)
127. Lissauer J J, Safronov V S *Icarus* **93** 288 (1991)
128. Black D C, Matthews M S (Eds) *Protostars and Planets II* (Tucson, AZ: Univ. Arizona Press, 1985)
129. Safronov V S *Icarus* **94** 260 (1991)
130. Stern S A *Icarus* **90** 271 (1991)
131. Weidenschilling S J *Icarus* **22** 426 (1974)
132. Beaugé C, Aarseth S J *Mon. Not. R. Astron. Soc.* **245** 30 (1990)
133. Chambers J E *Icarus* **152** 205 (2001)
134. Chambers J *Icarus* **180** 496 (2006)
135. Chambers J E *Icarus* **224** 43 (2013)
136. Chambers J E, Wetherill G W *Icarus* **136** 304 (1998)
137. Cox L P, Lewis J S *Icarus* **44** 706 (1980)
138. Dole S H *Icarus* **13** 494 (1970)
139. Hansen B M S *Astrophys. J.* **703** 1131 (2009)
140. Hoffmann V et al. *Mon. Not. R. Astron. Soc.* **465** 2170 (2017)
141. Ipatov S I *Sov. Astron.* **25** 617 (1981); *Astron. Zh.* **58** 1085 (1981)
142. Ipatov S I, Preprint No. 144 (Moscow: Institute of Applied Mathematics of the Academy of Sciences of the USSR, 1982)
143. Ipatov S I *Solar Syst. Res.* **21** 129 (1987); *Astron. Vestn.* **21** 207 (1987)
144. Ipatov S I *Int. Appl. Mech.* **28** 771 (1992); *Priklad. Mekh.* **28** (11) 91 (1992)
145. Ipatov S I *Solar Syst. Res.* **27** 65 (1993); *Astron. Vestn.* **27** 83 (1993)
146. Ipatov S I *Solar Syst. Res.* **53** 332 (2019); *Astron. Vestn.* **53** 349 (2019); arXiv:2003.11301
147. Izidoro A et al. *Astrophys. J.* **782** 31 (2014)
148. Lecar M, Aarseth S J *Astrophys. J.* **305** 564 (1986)
149. Lykawka P S, Ito T *Astrophys. J.* **838** 106 (2017)
150. Morbidelli A et al. *Annu. Rev. Earth Planet. Sci.* **40** 251 (2012); arXiv:1208.4694
151. Morishima R, Stadel J, Moore B *Icarus* **207** 517 (2010)
152. O'Brien D P, Morbidelli A, Levison H F *Icarus* **184** 39 (2006)
153. Raymond S N, Quinn T, Lunine J I *Icarus* **168** 1 (2004)
154. Raymond S N, Quinn T, Lunine J I *Icarus* **183** 265 (2006)
155. Raymond S N et al. *Icarus* **203** 644 (2009)
156. Wetherill G W *Science* **228** 877 (1985)
157. Wetherill G W, in *Mercury* (Eds F Vilas, C Chapman, M S Matthews) (Tucson, AZ: Univ. of Arizona Press, 1988) p. 670
158. Wetherill G W, in *Origins and Extinctions* (Eds D O Osterbrock, P H Raven) (New Haven, CT: Yale Univ. Press, 1988) p. 43
159. Wetherill G W, Stewart G R *Icarus* **77** 330 (1989)
160. Kozlov N N, Eneev T M, Preprint No. 134 (Moscow: Institute of Applied Mathematics of the Academy of Sciences of the USSR, 1977)
161. Eneev T M, Kozlov N N *Astron. Lett.* **5** 252 (1979); *Pis'ma Astron. Zh.* **5** 470 (1979)
162. Eneev T M, Kozlov N N *Solar Syst. Res.* **15** 97 (1982); *Astron. Vestn.* **15** 131 (1981)
163. Eneev T M, Kozlov N N *Dokl. Akad. Nauk SSSR* **263** 820 (1982)
164. Ipatov S I, Preprint No. 101 (Moscow: Institute of Applied Mathematics of the Academy of Sciences of the USSR, 1978)
165. Cox L P "Numerical simulation of the final stages of terrestrial planet formation", PhD Thesis (Cambridge: Massachusetts Institute of Technology, 1978)
166. Ipatov S I *Mat. Modelirovanie* **5** (1) 35 (1993)
167. Walsh K J et al. *Nature* **475** 206 (2011)
168. Jacobson S A, Morbidelli A *Phil. Trans. R. Soc. A* **372** 20130174 (2014)
169. O'Brien D P et al. *Icarus* **239** 74 (2014)
170. Rubie D C et al. *Icarus* **248** 89 (2015)
171. Dorofeeva V A, Makalkin A B *Evolyutsiya Rannei Solnechnoi Sistemy. Kosmokhimicheskie i Fizicheskie Aspekty* (Evolution of the Early Solar System. Cosmochemical and Physical Aspects) (Moscow: URSS, 2004)
172. Morbidelli A et al. *Astron. J.* **140** 1391 (2010)
173. Clement M S et al. *Icarus* **311** 340 (2018)
174. Clement M S et al. *Icarus* **321** 778 (2019)
175. Zharkov V N, Kozenko A V *Sov. Astron. Lett.* **16** 73 (1990); *Pis'ma Astron. Zh.* **16** 169 (1990)
176. Ipatov S I *Sov. Astron. Lett.* **17** 113 (1991); *Pis'ma Astron. Zh.* **17** 269 (1991)
177. Bouvier L C et al. *Nature* **55** 586 (2018)
178. Mezger K, Debaille V, Kleine T *Space Sci. Rev.* **174** 27 (2013)
179. Nimmo F, Kleine T *Icarus* **191** 497 (2007)
180. Elkins-Tanton L T *Earth Planet. Sci. Lett.* **271** 181 (2008)
181. Elkins-Tanton L T *Nature* **558** 522 (2018)
182. Lammer H et al. *Icarus* **339** 113551 (2020)
183. Ohtsuki K, Nakagawa Y, Nakazawa K *Icarus* **75** 552 (1988)
184. Kokubo E, Ida S *Icarus* **143** 15 (2000)
185. Weidenschilling S J et al. *Icarus* **128** 429 (1997)
186. Marov M Ya, Ipatov S I *Solar Syst. Res.* **52** 392 (2018); *Astron. Vestn.* **52** 402 (2018); arXiv:2003.09982
187. Ruskol E L *Sov. Astron.* **4** 657 (1961); *Astron. Zh.* **37** 690 (1960)
188. Ruskol E L *Sov. Astron.* **15** 646 (1972); *Astron. Zh.* **48** 819 (1971)

189. Ruskol E L *Proiskhozhdenie Luny* (Origin of the Moon) (Moscow: Nauka, 1975)
190. Afanas'ev V N, Pechernikova G V *Solar Syst. Res.* **56** 382 (2002); *Astron. Vestn.* **56** 389 (2022)
191. Hartmann W K, Davis D R *Icarus* **24** 504 (1975)
192. Cameron A G W, Ward W R, in *Lunar and Planetary Science Conf., Abstracts* Vol. 7 (Houston, TX: Lunar Science Institute, 1976) p. 120
193. Canup R M *Icarus* **168** 433 (2004)
194. Canup R M *Science* **338** 1052 (2012)
195. Canup R M, Asphaug E *Nature* **412** 708 (2001)
196. Canup R M, Barr A C, Crawford D A *Icarus* **222** 200 (2013)
197. Ćuk M, Stewart S T *Science* **338** 1047 (2012)
198. Ćuk M et al. *Nature* **539** 402 (2016)
199. Barr A C *J. Geophys. Res. Planets* **121** 1573 (2016); arXiv:1608.08959
200. Galimov E M et al. *Geokhimiya* (11) 1139 (2005)
201. Galimov E M et al. *Geokhimiya* (6) 563 (2011)
202. Galimov E M, Krivtsov A M *Origin of the Moon. New Concept: Geochemistry and Dynamics* (Berlin: De Gruyter, 2012)
203. Elkins-Tanton L T *Nature Geosci.* **6** 996 (2013)
204. Clery D *Science* **342** 183 (2013)
205. Jones J H, in *Proc. Lunar and Planet. Sci. Origin of the Earth and Moon Conf., Monterey, CA, 1998*, #4045
206. Ringwood A E *Earth Planet. Sci. Lett.* **95** 208 (1989)
207. Vityazev A V, Pechernikova G V *Fiz. Zemli* (6) 3 (1996)
208. Gorkavyi N N *Bull. Am. Astron. Soc.* **36** 861 (2004)
209. Gor'kavyi N N *Izv. Krymskoi Astrofiz. Observ.* **103** 143 (2007)
210. Svetsov V V, Pechernikova G V, Vityazev A V, in *43rd Lunar and Planetary Science Conf., 2012*, #1808
211. Citron R I et al., in *45th Lunar and Planetary Science Conf., 2014*, #2085
212. Rufu R, Aharonson O, Perets H B *Nature Geosci.* **10** 89 (2017)
213. Galimov E M, in *Problemy Zarozhdeniya i Evolyutsii Biosfery* (Problems of the Origin and Evolution of the Biosphere) (Ed. E M Galimov) (Moscow: Nauka, 1995) p. 8
214. Galimov E M, in *Problemy Zarozhdeniya i Evolyutsii Biosfery* (Problems of the Origin and Evolution of the Biosphere) (Ed. E M Galimov) (Moscow: Librokom, 2008) p. 213
215. Galimov E M, in *Problemy Zarozhdeniya i Evolyutsii Biosfery* (Problems of the Origin and Evolution of the Biosphere) (Ed. E M Galimov) (Moscow: Krasand, 2013) p. 47
216. Vasil'ev S V, Krivtsov A M, Galimov E M *Solar Syst. Res.* **45** 410 (2011); *Astron. Vestn.* **45** 420 (2011)
217. Okabayashi S et al. *Geochim. Cosmochim. Acta* **269** 1 (2020)
218. Kleine T et al. *Nature* **418** 952 (2002)
219. Yin Q et al. *Nature* **418** 949 (2002)
220. Williams C D, Sujoy M *Nature* **565** 78 (2019)
221. Ipatov S I *Solar Syst. Res.* **52** 401 (2018); *Astron. Vestn.* **52** 411 (2018); arXiv:2003.09925
222. Prokof'eva V V, Tarashchuk V P, Gor'kavyi N N *Phys. Usp.* **38** 623 (1995); *Usp. Fiz. Nauk* **165** 661 (1995)
223. Ipatov S I *Mon. Not. R. Astron. Soc.* **403** 405 (2010); arXiv:0904.3529
224. Ipatov S I *Solar Syst. Res.* **51** 409 (2017); *Astron. Vestn.* **51** 441 (2017); arXiv:1801.05254
225. Ipatov S I *Solar Syst. Res.* **51** 294 (2017); *Astron. Vestn.* **51** 321 (2017); arXiv:1801.05217
226. Marov M Ya et al. *Formirovanie Luny i Rannyaya Evolyutsiya Zemli* (Formation of the Moon and Early Evolution of the Earth) (Moscow: URSS, 2019)
227. Barricelli N A, Aashamar K *The Moon and the Planets* **22** 385 (1980)
228. Ipatov S I, Preprint No. 117 (Moscow: Institute of Applied Mathematics of the Academy of Sciences of the USSR, 1983)
229. Ipatov S I, Preprint No. 1 (Moscow: Institute of Applied Mathematics of the Academy of Sciences of the USSR, 1984)
230. Ipatov S I *Earth Moon Planets* **39** 101 (1987)
231. Ipatov S I *Solar Syst. Res.* **23** 16 (1989); *Astron. Vestn.* **23** 27 (1989)
232. Ipatov S I *Solar Syst. Res.* **23** 119 (1989); *Astron. Vestn.* **23** 197 (1989)
233. Ip W-H *Icarus* **80** 167 (1989)
234. Fernandez J A, Ip W-H *Icarus* **47** 470 (1981)
235. Fernandez J A, Ip W-H *Icarus* **54** 377 (1983)
236. Fernandez J A, Ip W-H *Icarus* **58** 109 (1984)
237. Fernandez J A, Ip W-H *Planet. Space Sci.* **44** 431 (1996)
238. Zharkov V N, in *Evolution of the Earth and Planets* (Geophysical Monograph Ser., Vol. 74, Eds E Takahashi, R Jeanloz, D Rubie) (New York: John Wiley and Sons, 1993) p. 7
239. Ipatov S I, in *22nd Lunar and Planetary Science Conf., 1991*, p. 607
240. Thommes E W et al. *Nature* **402** 635 (1999)
241. Gomes R et al. *Nature* **435** 466 (2005)
242. Morbidelli A et al. *Nature* **435** 462 (2005)
243. Tsiganis K et al. *Nature* **435** 459 (2005)
244. Gudkova T V, Zharkov V N *Planet. Space Sci.* **47** 1201 (1999)
245. Zharkov V N *Solar Syst. Res.* **25** 465 (1991); *Astron. Vestn.* **25** 627 (1991)
246. Zharkov V N *Vnutrennee Stroenie Zemli i Planet. Elementarnoe Vvedenie v Planetnuyu i Sputnikovuyu Geofiziku* (Internal Structure of the Earth and Planets. The Elementary Introduction to Planetary and Satellite Geophysics) (Moscow: Nauka i Obrazovanie, 2013)
247. Zharkov V N, Gudkova T V *Izv. Phys. Solid Earth* **55** 50 (2019); *Fiz. Zemli* (1) 61 (2019)
248. Lodders K *Astrophys. J.* **591** 1220 (2003)
249. Ziglina I N, Makalkin A B *Solar Syst. Res.* **50** 408 (2016); *Astron. Vestn.* **50** 431 (2016)
250. Makalkin A B, Artyushkova M E *Solar Syst. Res.* **51** 491 (2017); *Astron. Vestn.* **51** 524 (2017)
251. D'Angelo G, Lissauer J J, in *Handbook of Exoplanets* (Eds H Deeg, J Belmonte) (Cham: Springer, 2018) p. 2319, https://doi.org/10.1007/978-3-319-55333-7_140; arXiv:1806.05649
252. Subbotin M F *Vvedenie v Teoreticheskuyu Astronomiyu* (Introduction to Theoretical Astronomy) (Moscow: Nauka, 1968)
253. Applegate J H et al. *Astron. J.* **92** 176 (1986)
254. Nobili A M, Milani A, Carpino M *Astron. Astrophys.* **210** 313 (1989)
255. Laskar J *Astron Astrophys.* **198** 341 (1988)
256. Ipatov S I *Solar Syst. Res.* **34** 179 (2000); *Astron. Vestn.* **34** 195 (2000)
257. Cameron A G W, Pine M R *Icarus* **18** 377 (1973)
258. Torbett M, Smoluchowski R *Icarus* **44** 722 (1980)
259. Wetherill G W, in *Asteroids II. Proc. of the Conf., Tucson, AZ, March 8–11, 1988* (Ed. R Binzel) (Tucson, AZ: Univ. of Arizona Press, 1989) p. 661
260. Raymond S N, Izidoro A *Icarus* **297** 134 (2017)
261. Stern S A *Astron. J.* **110** 856 (1995)
262. Stern S A *Astron. J.* **112** 1203 (1996)
263. Stern S A *Astron. Astrophys.* **310** 999 (1996)
264. Stern S A, in *Solar System Ices. Based on Reviews Presented at the Intern. Symp. "Solar System Ices", Toulouse, France, on March 27–30, 1995* (Astrophysics and Space Science Library, Vol. 227, Eds B Schmitt, C De Bergh, M Festou) (Dordrecht: Kluwer Acad. Publ., 1998) p. 685
265. Stern S A, Colwell J E *Astron. J.* **114** 841 (1997)
266. Kenyon S J, Luu J X *Astron. J.* **115** 2136 (1998)
267. Kenyon S J, Luu J X *Astron. J.* **118** 1101 (1999)
268. Davis D R, Farinella P *Icarus* **125** 50 (1997)
269. Durda D D, Stern S A *Icarus* **145** 220 (2000)
270. Eneev T M *Sov. Astron. Lett.* **6** 163 (1980); *Pis'ma Astron. Zh.* **6** 295 (1980)
271. Ipatov S I *Adv. Space Res.* **28** 1107 (2001)
272. Ipatov S I, Preprint No. 43 (Moscow: Institute of Applied Mathematics of the Academy of Sciences of the USSR, 1980)
273. Ipatov S I *Kinematics Phys. Celest. Bodies* **4** (6) 76 (1988); *Kinematika Fiz. Nebesnykh Tel* **4** (6) 73 (1988)
274. Ipatov S I, Henrard J *Solar Syst. Res.* **34** 61 (2000); *Astron. Vestn.* **34** 68 (2000)
275. Petit J-M, Morbidelli A, Valsecchi G B *Icarus* **141** 367 (1999)
276. Marov M Ya, in *Nasledie Galileya. Sbornik Lektsii, Prochitannykh na Konf. "Astronomiya i Obshchestvo", Moscow, 25–27 Marta 2009 Goda* (Galileo's Legacy. Collection of Lectures Read at the Conf. "Astronomy and Society", Moscow, March 25–27, 2009) (Eds A A Boyarchuk, D Z Vibe) (Moscow: Skif, 2009) p. 108
277. Genda H, Icoma M *Icarus* **194** 42 (2008)
278. Drake M J, Campins H, in *Asteroids, Comets, Meteors, Proc. of the 229th Symp. of the Intern. Astronomical Union, Buzios, Rio de Janeiro, Brasil, August 7–12, 2005* (Eds D Lazzaro, S Ferraz-Mello, J A Fernández) (Cambridge: Cambridge Univ. Press, 2006) p. 381
279. Muralidharan K et al. *Icarus* **198** 400 (2008)

30    M Ya Marov, S I Ipatov

280. Sobolev A V et al. *Nature* **531** 628 (2016)
281. Hallis L J et al. *Science* **350** 795 (2015)
282. Morbidelli A et al. *Meteorit. Planet. Sci.* **35** 1309 (2000)
283. Petit J-M, Morbidelli A, Chambers J *Icarus* **153** 338 (2001)
284. Raymond S N, Quinn T, Lunine J I *Icarus* **168** 1 (2004)
285. Lunine J I et al. *Icarus* **165** 1 (2003)
286. Lunine J et al., in *38th Lunar and Planetary Science Conf.*, 2007, #1616
287. Levison H F et al. *Icarus* **151** 286 (2001)
288. Marov M Ya, Ipatov S I, in *Collisional Processes in the Solar System* (Astrophysics and Space Science Library, Vol. 261, Eds M Ya Marov, H Rickman) (Dordrecht: Kluwer Acad. Publ., 2001) p. 223
289. Marov M Ya, Ipatov S I *Solar Syst. Res.* **39** 374 (2005); *Astron. Vestn.* **39** 419 (2005)
290. Ipatov S I, Mather J C *Earth, Moon, Planets* **92** 89 (2003); astro-ph/0305519
291. Ipatov S I, Mather J C *Adv. Space Res.* **33** 1524 (2004); astro-ph/0212177
292. Ipatov S I, Mather J C *Adv. Space Res.* **37** 126 (2006); astro-ph/0411004
293. Ipatov S I, Mather J C, in *Near-Earth Objects, our Celestial Neighbors: Opportunity and Risk* (Proc. of the IAU Symp., No 236, Eds A Milani, G B Valsecchi, D Vokrouhlický) (Cambridge: Cambridge Univ. Press, 2007) p. 55; astro-ph/0609721
294. Ipatov S I, in *Icy Bodies in the Solar System* (Proc. of the IAU Symp., No 263, Eds J A Fernandez et al.) (Cambridge: Cambridge Univ. Press, 2010) p. 41; arXiv:0910.3017
295. Dorofeeva V A *Solar Syst. Res.* **54** 96 (2020); *Astron. Vestn.* **54** 110 (2020)
296. Delsemme A H *Planet. Space Sci.* **47** 125 (1999)
297. Yang L, Ciesla F J, Alexander C M O'D *Icarus* **226** 256 (2013)
298. Zheng X, Lin D N C, Kouwenhoven M B N *Astrophys. J.* **835** 207 (2017)
299. Pavlov A A, Pavlov A K, Kasting J F *J. Geophys. Res.* **104** 30725 (1999)
300. Levison H F, Duncan M *Icarus* **108** 18 (1994)
301. Ipatov S I *Solar Syst. Res.* **29** 261 (1995); *Astron. Vestn.* **29** 304 (1995)
302. Marov M Ya, Grinspoon D H *The Planet Venus* (New Haven, CT: Yale Univ. Press, 1998)
303. Shustov B M *Phys. Usp.* **54** 1068 (2011); *Usp. Fiz. Nauk* **181** 1104 (2011)
304. Emel'yanenko V V, Shustov B M *Phys. Usp.* **56** 833 (2013); *Usp. Fiz. Nauk* **183** 885 (2013)
305. Beskin G M et al. *Phys. Usp.* **56** 836 (2013); *Usp. Fiz. Nauk* **183** 888 (2013)
306. Shustov B M et al. *Astron. Rep.* **59** 983 (2015); *Astron. Zh.* **92** 867 (2015)
307. Marov M Ya, in *Astrophysics. New Research* (Ed. C Lloyd) (New York: Nova Publ., 2017) p. 27
308. Bottke W F (Jr.) et al. *Icarus* **156** 399 (2002)
309. Shustov B M, Rykhlova L V (Eds) *Asteroidno-Kometnaya Hazard: Vchera, Segodnya, Zavtra* (Asteroid-Comet Danger: Yesterday, Today, Tomorrow) (Moscow: Fizmatlit, 2010)
310. Popova O P et al. *Science* **242** 1069 (2013)
311. Farinella P et al. *Icarus* **101** 174 (1993)
312. Eneev T M, Preprint No. 166 (Moscow: Institute of Applied Mathematics of the Academy of Sciences of the USSR, 1979)
313. Wetherill G W *Icarus* **76** 1 (1988)
314. Öpik E J *Adv. Astron. Astrophys.* **2** 219 (1963)
315. Britt D T et al. *Icarus* **99** 153 (1992)
316. Shoemarker E M, Weissman P R, Shoemaker C S, in *Hazards Due to Comets and Asteroids* (Space Science Ser., Eds T Gehrels, M S Matthews, A M Schumann) (Tucson, AZ: Univ. of Arizona Press, 1994) p. 313
317. Morbidelli A et al. *Astron. Astrophys.* **282** 955 (1994)
318. Ferraz-Mello S, in *Asteroids, Comets, Meteors 1993: Proc. of the 160th Symp. of the Intern. Astronomical Union, Belgirate, Italy, June 14–18, 1993* (Intern. Astronomical Union. Symp., No. 160, Eds A Milani, M DiMartino, C Cellino) (Dordrecht: Kluwer Acad. Publ., 1994) p. 175
319. Froeschle Ch, Scholl H *Celest. Mech. Dyn. Astron.* **46** 231 (1989)
320. Froeschle Ch, Morbidelli A, Scholl H *Astron. Astrophys.* **249** 553 (1991)
321. Scholl H, Froeschle C F *Astron. Astrophys.* **227** 255 (1990)
322. Scholl H, Froeschle C F *Astron. Astrophys.* **245** 316 (1991)
323. Wetherill G W *Science* **228** 877 (1985)
324. Ipatov S I *Sov. Astron. Lett.* **15** 324 (1989); *Pis'ma Astron. Zh.* **15** 750 (1989)
325. Ipatov S I *Solar Syst. Res.* **26** 520 (1992); *Astron. Vestn.* **26** 26 (1992)
326. Ipatov S I *Icarus* **95** 100 (1992)
327. Wisdom J *Astron. J.* **87** 577 (1982)
328. Wisdom J *Icarus* **56** 51 (1983)
329. Ipatov S I *Kinematics Phys. Celest. Bodies* **4** (4) 49 (1988); *Kinematika Fiz. Nebesnykh Tel* **4** (4) 49 (1988)
330. Sidlichovsky M *Publ. Astron. Inst. Czechosl. Acad. Sci.* **68** 33 (1987)
331. Ipatov S I, Preprint No. 30 (Moscow: Institute of Applied Mathematics of the Academy of Sciences of the USSR, 1980)
332. Greenberg R, Scholl H, in *Asteroids* (Ed. T Gehrels) (Tucson, AZ: Univ. of Arizona Press, 1979) p. 310
333. Yoshikawa M *Astron. Astrophys.* **213** 436 (1989)
334. Yoshikawa M *Icarus* **92** 94 (1991)
335. Hahn G et al. *Astron. Astrophys.* **246** 603 (1991)
336. Scholl H, Froeschle C F *Astron. Astrophys.* **42** 457 (1975)
337. Wisdom J *Icarus* **56** 51 (1983)
338. Yoshikawa M *Icarus* **87** 78 (1990)
339. Migliorini F et al. *Science* **281** 2022 (1998)
340. Morbidelli A, Nesvorny D *Icarus* **139** 295 (1999)
341. Gladman B, Michel P, Froeschlé Ch *Icarus* **146** 176 (2000)
342. Fernández J A *Mon. Not. R. Astron. Soc.* **192** 481 (1980)
343. Duncan M J, Levison H F, Budd S M *Astron. J.* **110** 3073 (1995)
344. Kuchner M J, Brown M E, Holman M *Astron. J.* **124** 1221 (2002)
345. Kazimirchak-Polonskaya E I *Byull. Inst. Teor. Astron.* **12** 796 (1971)
346. Levison H F, Duncan M J *Icarus* **127** 13 (1997)
347. Harris N W, Bailey M E *Mon. Not. R. Astron. Soc.* **297** 1227 (1998)
348. Levin B Yu *Sov. Phys. Usp.* **8** 360 (1965); *Usp. Fiz. Nauk* **86** 41 (1965)
349. Jewitt D, Luu J, Chen J *Astron. J.* **112** 1225 (1996)
350. Pitjeva E V, Pitjev N P *Astron. Lett.* **44** 544 (2018); *Pis'ma Astron. Zh.* **44** 604 (2018)
351. Jewitt D, Fernandez Y, in *Collisional Processes in the Solar System* (Astrophysics and Space Science Library, Volume 261, Eds M Ya Marov, H Rickman) (Dordrecht: Kluwer Acad. Publ., 2001) p. 143
352. Jewitt D *Annu. Rev. Earth. Planet. Sci.* **27** 287 (1999)
353. Luu J et al. *Nature* **387** 573 (1997)
354. Trujillo C A, Jewitt D C, Luu J X *Astrophys. J.* **529** L103 (2000)
355. Ipatov S I, Hahn G J *Solar Syst. Res.* **33** 487 (1999); *Astron. Vestn.* **33** 553 (1999)
356. Ipatov S I *Celest. Mech. Dyn. Astronomy* **73** 107 (1999)
357. Bailey M E, Emel'yanenko V V *Mon. Not. R. Astron. Soc.* **278** 1087 (1996)
358. Emel'yanenko V V *Celest. Mech. Dyn. Astron.* **54** 91 (1992)
359. Marov M Ya, Ipatov S I *Geochem. Int.* **59** 1010 (2021); *Geokhimiya* **66** 964 (2021)
360. Nakamura T, Kurahashi H *Astron. J.* **115** 848 (1998)
361. Fortov V E et al. *Phys. Usp.* **39** 363 (1996); *Usp. Fiz. Nauk* **166** 391 (1966)
362. Ksanfomality L V *Phys. Usp.* **55** 137 (2012); *Usp. Fiz. Nauk* **182** 147 (2012)
363. Hartmann W K *Meteoritics* **30** 451 (1995)
364. Werner S C, Ivanov B A *Treatise Geophys. Second Ed.* **10** 327 (2015)
365. Bottke W F (Jr.) et al., in *Hazards Due to Comets and Asteroids* (Ed. T Gehrels, M S Matthews, A Schumann) (Tucson, AZ: Univ. of Arizona Press, 1994) p. 337
366. Shoemaker E M, Wolf R F, Shoemaker C S, in *Global Catastrophes in Earth History: an Interdisciplinary Conf. on Impacts, Volcanism, and Mass Mortality* (Special Paper, Geological Society of America, 247, Eds V L Sharpton, P D Ward) (Boulder, CO: Geological Soc. of America, 1990) p. 155
367. Bottke W, Vokrouhlicky D, Nesvorny D *Nature* **449** 48 (2007)
368. Mazrouei S et al. *Science* **363** 253 (2019)
369. McEwen A S, Moore J M, Shoemaker E M *J. Geophys. Res. Planets* **102** 9231 (1997)
370. Ipatov S I, Feoktistova E A, Svetsov V V *Solar Syst. Res.* **54** 384 (2020); *Astron. Vestn.* **54** 409 (2020); arXiv:2011.00361

371. Losiak A et al. "Lunar Impact Crater Database", Lunar Exploration Intern. Program (Houston, TX: Lunar and Planetary Inst., 2009) updated by T Ohman Planet. Space Sci., 2015
372. Feoktistova E A, Ipatov S I *Earth Moon Planets* **125** 1 (2021); arXiv:2103.00291
373. Whipple F L *IAU Circ.* (3881) 1 (1983)
374. Drolshagena G et al. *Planet. Space Sci.* **143** 21 (2017)
375. Gor'kavyi N N, Ozernoy L M, Mather J C *Astrophys. J.* **474** 496 (1997)
376. Liou J-C, Dermott S F, Xu Y L *Planet. Space Sci.* **43** 717 (1995)
377. Liou J-C, Zook H A, Dermott S F *Icarus* **124** 429 (1996)
378. Moro-Martin A, Malhotra R *Astron. J.* **124** 2305 (2002)
379. Moro-Martin A, Malhotra R *Astron. J.* **125** 2255 (2003)
380. Nesvorný D et al. *Icarus* **181** 107 (2006)
381. Reach W T, Franz B A, Weiland J L *Icarus* **127** 461 (1997)
382. Trofimov P M, Gorkavyi N N *Solar Syst. Res.* **56** 183 (2022); *Astron. Vestn.* **56** 198 (2022)
383. Gor'kavyi N N, Trofimov P M *Solar Syst. Res.* **56** 225 (2022); *Astron. Vestn.* **56** 237 (2022)
384. Ipatov S I, Mather J C, Taylor P A *Ann. New York Acad. Sci.* **1017** 66 (2004); astro-ph/0308450
385. Ipatov S I, in *The Ninth Moscow Solar System Symp. 9M-S3, Moscow, Russia, October 8–12, 2018*, #9MS3-DP-01, p. 144; https://ms2018.cosmos.ru
386. Ipatov S I et al. *Icarus* **194** 769 (2008); arXiv:0711.3494
387. Reynolds R J, Madsen G J, Moseley S H *Astrophys. J.* **612** 1206 (2004)
388. Jorgensen J L et al. *J. Geophys. Res. Planets* **126** e2020JE006509 (2020)
389. Schwarz R et al. *Mon. Not. R. Astron. Soc.* **480** 3595 (2018)
390. Ipatov S I *Bull. Am. Astron. Soc.* **53** 2021n3i1126 (2021)
391. Ipatov S I, in *Thirteenth Moscow Solar System Symp., 13M-S3, 2022* (13MS3-EP-08) p. 372
392. Kervella P, Arenou F, Schneider J *Astron. Astrophys.* **635** L14 (2020)
393. Benedict G F, McArthur B E *Res. Notes AAS* **4** 86 (2020)
394. Ipatov S, EPSC Abstracts, Vol.14, EPSC2020-71, 2020, updated on 08 Oct 2020 (2020) https://doi.org/10.5194/epsc2020-71
395. Canup R M, Pierazzo E, in *37th Lunar Planetary Science Conf., 2006*, #2146
396. Ciesla F J et al. *Astrophys. J.* **804** 9 (2015)
397. Fulle M et al. *Mon. Not. R. Astron. Soc.* **469** S45 (2017)
398. Raymond S N et al. *Nat. Astron.* **6** 80 (2022)
399. Ipatov S I *Solar Syst. Res.* **57** (3) (2023); *Astron. Vestn.* **57** (3) (2023), in print p. 236
400. Ipatov S I *Meteorit. Planet. Sci.* **57** (S1) A208 (2022)